\documentclass[journal,twocolumn,10pt]{IEEEtran}
\IEEEoverridecommandlockouts 
\usepackage{epsfig,latexsym}
\usepackage{float}
\usepackage{indentfirst}
\usepackage{amsmath}
\usepackage{amssymb}
\usepackage{times}
\usepackage{graphicx}
\usepackage{subfigure}
\usepackage{pifont}
\usepackage{psfrag}
\usepackage{cite}
\usepackage{url}
\usepackage{lastpage}
\usepackage{epstopdf}
\usepackage{bm}
\usepackage{color}
\usepackage{cases}
\usepackage{fancyhdr}
\usepackage[center]{caption2}
\usepackage{balance}
\usepackage{fancyhdr,lastpage}
\begin{document}
\title{Secure Wireless Powered and Cooperative Jamming D2D Communications}
\author{
Zheng Chu,~\IEEEmembership{Member,~IEEE}, Huan X. Nguyen,~\IEEEmembership{Senior Member,~IEEE} \\Tuan Anh Le,~\IEEEmembership{Member,~IEEE}, Mehmet Karamanoglu,~\IEEEmembership{Member,~IEEE}\\Enver Ever,~\IEEEmembership{Member,~IEEE}, and Adnan Yazici,~\IEEEmembership{Senior Member,~IEEE}
\vspace{-0.3in}
\thanks{Part of this work has been presented in Wireless Days (WD) Conference, Porto, Portugal, 2017 \cite{Zheng_Chu_WD_2017}.}
\thanks{This work was supported by the Newton Fund/British Council Institutional Links under Grant ID 216429427, Project code 101977.}
\thanks{Z. Chu, H. Nguyen, T. Le, and M. Karamanoglu are with the Faculty of Science and Technology, Middlesex University, The Burroughs, London NW4 4BT, U.K. (Email: z.chu@mdx.ac.uk; h.nguyen@mdx.ac.uk; t.le@mdx.ac.uk; M.Karamanoglu@mdx.ac.uk)}
\thanks{E. Ever, and A. Yazici are with the Department of Computer Engineering, Middle East Technical University, Turkey. (Email: eever@metu.edu.tr; yazici@ceng.metu.edu.tr).}
}
\vspace{-0.2in}
\maketitle
\thispagestyle{empty}
\begin{abstract}
This paper investigates a secure wireless-powered device-to-device (D2D) communication network in the presence of multiple eavesdroppers, where a hybrid base station (BS) in a cellular network not only provides power wirelessly for the D2D transmitter to guarantee power efficiency for the D2D network, but also serves as a cooperative jammer (CJ) to interfere with the eavesdroppers.
The cellular and D2D networks can belong to different service providers, which means that the D2D transmitter would need to pay for the energy service released by the hybrid BS to guarantee secure D2D communication. In order to exploit the hierarchical interaction between the BS and the D2D transmitter, we first formulate a \emph{Stackelberg} game based energy trading scheme, where the quadratic energy cost model is considered. Then, a non-energy trading based \emph{Stackelberg} game is investigated to study the reversed roles of the BS and the D2D users. For comparison, we also formulate and resolve the social welfare optimization problem.
We derive the closed-form \emph{Stackelberg} equilibriums of the formulated games and the optimal solutions for the social welfare optimization problem. Simulation results are provided to validate our proposed schemes to highlight the importance of energy trading interaction between cellular and D2D networks.
\end{abstract}
\IEEEpeerreviewmaketitle
\setlength{\baselineskip}{1\baselineskip}
\newtheorem{definition}{Definition}
\newtheorem{fact}{Fact}
\newtheorem{assumption}{Assumption}
\newtheorem{theorem}{Theorem}
\newtheorem{lemma}{Lemma}
\newtheorem{corollary}{Corollary}
\newtheorem{proposition}{Proposition}
\newtheorem{example}{Example}
\newtheorem{remark}{Remark}
\newtheorem{algorithm}{Algorithm}
\section{Introduction}
Device-to-device (D2D) communications, which was standardized by the 3GPP release 12
\cite{Asadi_D2D_IEEE_Com_Sur_Tuts_2014,Andrews_D2D_3GPP_IEEE_CM_2014}, has been recently receiving increasing research interests as one of the driving technologies for 5G networks.
The key feature of D2D communication is that two
communicating devices in close proximity reuse better
links to communicate directly rather than through a base station (BS) in cellular networks \cite{Lingyang_Song_Game_D2D_IEEEWC_2014}.
The D2D communication based mobile proximity offers new mobile service opportunities and the potential to reduce traffic load on the network. The advantages of the D2D communications are multi-fold: relieving the burden of
the BS, enhancing spectral efficiency, shortening time
delay, and reducing power consumption to keep up
with the ``\emph{greener}'' trend. D2D communication is also supposed to serve well in the urgent scenarios for providing public safety and disaster relief services \cite{Ali_D2D_Disaster_2015,Ali_D2D_PSNU_WCNC_2016,Zheng_ICT_2017}.
Recent information-theoretical results have indicated that the combination of caching on the users' devices and D2D communication leads to throughput scalability for very dense networks, see e.g., \cite{Caire_ToN_2016} and references therein. Hence, D2D caching networks have been considered as one of promising paradigms in \cite{Caire_ToN_2016,Caire_ToC_2015}, where it is advocated that ``helper" nodes are densely disseminated in a coverage area to serve the users' demands using their own large cached information. However, the works in \cite{Caire_ToN_2016} and \cite{Caire_ToC_2015} have implicitly assumed infinite power supply for helper nodes. Unfortunately, the infinite-power-supply assumption is not always practical for D2D transceivers.

To tackle the aforementioned problem, energy harvesting D2D networks have been considered as a promising solution to provide energy for the D2D transceivers \cite{Yan_Zhang_Network_2015}. Since energy can be harvested from ambient radio frequency (RF) signals with reasonable efficiency over small distances, RF energy harvesting can be adopted for D2D communications. Therefore, wireless energy transfer (WET) techniques have attracted increasing research interests to prolong the battery lifetime of energy-constrained wireless networks \cite{Varshney_08,Shannon_Tesla_10}. In particular, research efforts have been focusing on the establishment of wireless-powered communication networks (WPCNs), in which wireless transceivers are wirelessly charged by the power transmitters \cite{Rui_Zhang_WPCN_CM_2015,Rui_Zhang_WPCN_CM_2016}. Unlike traditional battery-powered networks, WPCN can effectively reduce the operational cost by avoiding the need to replace or recharge batteries. Also, the energy released by the WPCN is adjustable in providing a stable energy supply to satisfy different physical conditions and service requirements \cite{Rui_Zhang_WPCN_CM_2016}. In WPCNs, a well-known protocol, named ``\emph{harvest-then-transmit}'', was proposed in \cite{Rui_Zhang_WPCN_TWC_2014}, where wireless users first harvest energy from RF signals broadcast by a hybrid access-point (AP) in the downlink (DL), and then use the harvested energy to send information to the AP in the uplink (UL). State-of-the-art cooperative protocols for WPCNs can be found in \cite{Rui_Zhang_GLOBECOM_2014_UC_WPCN,He_Chen_ITW_WPCN_Relay_2014,He_Chen_TSP_HTC_WPCN_2016}. The authors of \cite{Kaibin_Huang_TWC_2014_WPT} and \cite{Kaibin_Huang_CM_2015} proposed a dedicated WET network where multiple power beacons (PBs) are deployed near a wireless information transfer (WIT) network to provide energy services to the wireless terminals, i.e., D2D transceivers, of the WIT network. 

On the other development, physical layer security, which differs from the conventional cryptographic methods developed in the network layer, has recently attracted significant research efforts, see e.g., \cite{Wyner_J75,Korner_Info_Theory_J78,Wornell_Info_Theory_J10,Wornell_Info_Theory1_J10}. Physical layer security has been initially defined in wiretap channel based on information-theoretical aspects \cite{Wyner_J75,Korner_Info_Theory_J78}. In physical layer security, a better channel of a legitimate user, compared with that of an eavesdropper, is exploited to guarantee secure transmissions for the legitimate user, i.e., to guarantee a positive secrecy rate which is defined as the mutual information difference between the legitimate user and the eavesdropper. To improve the secrecy rate, multi-antenna wiretap channels have been investigated to take the advantage of having the additional degrees of freedom (DoF) and diversity gains \cite{Wornell_Info_Theory_J10,Wornell_Info_Theory1_J10}. Most of existing techniques aim to introduce more interference to degrade the channels of eavesdroppers, i.e., artificial noise (AN) \cite{Ma_TSP_J13,TuanCL2015,TuanTGCN17} and cooperative jammer (CJ) \cite{Zheng_Secrecy_J15}. In \cite{Ma_Sig_Process_J11}, a semidefinite programming (SDP) approach was adopted for multiple-input single-output (MISO) secure channels to guarantee reliable communications by solving  either power minimization subject to secrecy constraint or secrecy rate maximization subject to power constraint. An AN-assisted transmit optimization has been presented in \cite{Ma_TSP_J13}, where the spatially selective AN embedded with secure transmit beamformer was designed to obtain the optimal power allocation. In \cite{Zheng_Secrecy_J15}, a scheme that utilizes a CJ signal sent by an external node is considered. The CJ signal is used to introduce interference to the eavesdroppers in order to achieve the required achievable secrecy rate for a multiple-input multiple-output (MIMO) secrecy channel. In \cite{Zhu_Han_Stackelberg_PLS_2010,Zheng_Secrecy_J15,Zheng_Stackelberg_game_EUSIPCO_2014}, several designs were proposed with an assumption that a legitimate transmitter pays a price to a private CJ for its jamming services to guarantee the secure communications. A \emph{Stackelberg} game was employed to obtain the optimal power allocation in \cite{Zheng_Secrecy_J15,Zheng_Stackelberg_game_EUSIPCO_2014}. In addition, outage probability is a key indicator to show secrecy performance when only statistical channel state information (CSI) is available. In \cite{Zheng_WCL_2015,Zheng_Sec_TWC_2016,TuanTGCN17}, outage secrecy rate optimizations were formulated based on the uncertainty model of statistical channels, where \emph{Bernstein-type} inequality \cite{Bernstein-type} is adopted to solve the formulated problems.

As an emerging application of D2D communications, a new cellular network architecture is proposed to integrate energy harvesting technologies and social networking characteristics into D2D communications for local data dissemination in \cite{Yan_Zhang_IEEEWC_D2D_EH_2016}. In D2D caching and social networks, the need for secrecy protection arises as on-demand broadcasting normally requires a subscription, which means unsubscribed users will not be able to access the content that they do not pay for. Thus, security communications should be considered in the D2D networks. To that end, physical layer security techniques have been adopted in D2D communications, where secure communications can be guaranteed such that confidential information is reliably transmitted between a dedicated D2D pair  \cite{Rongqi_Zhang_GLOBE_2015_D2D_PLS,Rongqi_Zhang_TWC_2016_D2D_PLS,Rongqing_Zhang_PLS_D2D_ICC_2016,Zheng_Sec_D2D_IET_COM_2015}. In \cite{Rongqi_Zhang_GLOBE_2015_D2D_PLS,Rongqi_Zhang_TWC_2016_D2D_PLS}, the authors proposed a novel cooperation mechanism between the cellular and D2D networks, which was formulated as a coalition game. In addition, a merge-and-split-based coalition formation algorithm is proposed to achieve efficient cooperation such that one can improve the secrecy rate and social welfare. In \cite{Rongqing_Zhang_PLS_D2D_ICC_2016}, the optimal joint power control solutions of both the cellular communication links
and D2D pairs have been derived in terms of secrecy capacity. In \cite{Zheng_Sec_D2D_IET_COM_2015}, the D2D network plays a CJ role to interfere with the eavesdroppers for the improvement of the security in cellular networks. On the other hand, the power signal in a dedicated WET for transferring wireless power, can be employed to guarantee the secrecy communications in WIT such that physical layer security techniques can be naturally applied to the secure WPCNs \cite{Xiaoming_Chen_CL_Sec_WPCN_2016}. The interaction between WET and security services plays an important role to the performance of the network, particularly when the WET supplier and the WIT users belong to different service providers.  To the best of our knowledge, there has been no existing work investigating this energy interaction in secure wireless-powered D2D communications. 

In this paper, we investigate a secure wireless-powered CJ-aided D2D communication system, where a hybrid BS plays two roles: WET and CJ-aided secure WIT. In addition, the D2D secure communication is established by using the harvested energy. The WET and CJ-aided secure WIT are performed in two phases: i) the BS is considered as a power source (PS) which transfers the power to the D2D transmitter in the first phase, and ii) the D2D transmitter employs the harvested energy to transmit the information signal while the BS acts as a CJ to enhance the secrecy performance.
Unlike the existing work \cite{Kaibin_Huang_TWC_2014_WPT}, where the WET and WIT networks are assumed to belong to the same service provider, in this paper, we consider a more general and complete case where these two networks may or may not belong to the same operators.\footnote{In the case of different operators, we assume that the WET service is provided by one service provider, e.g., an energy supplier, while the WIT service is provided by another provider, e.g., a  telecommunication supplier. Both of service providers belong to different authorities.} In this case, energy prices will be paid by the D2D transmitter for the wireless charging services to guarantee the secure transmission. We propose three different schemes to exploit the energy interaction between the BS and the D2D transmitter to achieve optimal power allocation for the cellular network while guaranteeing security for D2D communications. In the following, we highlight our contributions:
\begin{enumerate}
	\item \emph{Energy trading based \emph{Stackelberg} game}: We first consider the wireless charging model as an \emph{energy trading} process, which is defined as the hierarchical energy interaction between the BS and the D2D transmitter. This model facilitates the derivation of the optimal power allocation policies for the D2D devices and the BS. In particular, we take into account strategic behaviours of the hybrid BS and the D2D transmitter and formulate this \emph{energy trading} process as a \emph{Stackelberg} game. The D2D transmitter plays the leader role\footnote{As a customer, it can dictate the market by deciding what energy price that it is willing to pay and therefore it fits well with the role of the leader in this trading game.} that purchases energy from the BS's wireless power transfer service. In this game, the leader optimizes the energy price that it needs to pay and the amount of time allocated for energy harvesting to maximize its utility function, which is defined as the difference between the achievable secrecy throughput, i.e., the equivalent revenue, and its total payment to the BS. Meanwhile, the BS is modelled as the follower that decides its optimal power transfer policy based on the energy price offered by the leader to maximize its own profit, which is defined as the difference between the payment received from the D2D transmitter and the energy production cost. The \emph{Stackelberg} equilibrium of the game is then achieved via closed form solutions (for the BS's transmit power policy and the D2D transmitter's energy price) or via numerical search (for time allocation of the energy harvesting).
	\item \emph{Non-energy trading based Stackelberg game}: Next, we consider a non-energy trading based game-theoretical scheme. The nature of non-trading strategy is that the energy seller, i.e, the BS, decides the energy price and the amount of time for energy transfer service. There is no room for the customer, i.e., the D2D transmitter, to negotiate the price. To reflect this, the BS now plays the leader role, who dictates the situation, in the \emph{Stackelberg} game. The D2D transmitter is now modelled as the follower who determines the level of power transfer required from the BS. The closed form solutions are achieved for the optimal energy price and transmit power in this game.	
	\item \emph{Social welfare optimization scheme}: Finally, we formulate and solve a corresponding social welfare optimization scheme in order to evaluate the impact of the self-interested behaviours of the hybrid BS that is present in the \emph{energy trading} based \emph{Stackelberg} game. In this scheme, the hybrid BS is considered to cooperate with the D2D transmitter to maximize the social welfare which is defined as the difference between the utility function achieved from the achievable secrecy throughput of the D2D transmitter and the BS's energy cost presented by a quadratic model. We derive the closed-form solution for the BS transmit power allocation, while the optimal energy transfer time allocation can be achieved by numerical search.
\end{enumerate}

The rest of of the paper is organized as follows. Section \ref{section System model} presents the secure wireless-powered D2D system model. Section \ref{section Game_theoretical_scheme} proposes the game theoretical and social welfare optimization schemes. Numerical results are provided to validate the proposed schemes in Section \ref{section Numerical_results}. Finally, Section \ref{section:Conclusions} concludes the paper.

\textbf{\emph{Notations}:}
We use the upper case boldface letters for matrices and lower case boldface letters for vectors.
 $ \|\cdot\| $ denotes the Euclidean norm of a vector.
$[x]^{+}$ represents $\max\{x,0\}$.
\section{System Model}\label{section System model}
\begin{figure}[!htbp]
	\centering
	\includegraphics[scale = 0.33]{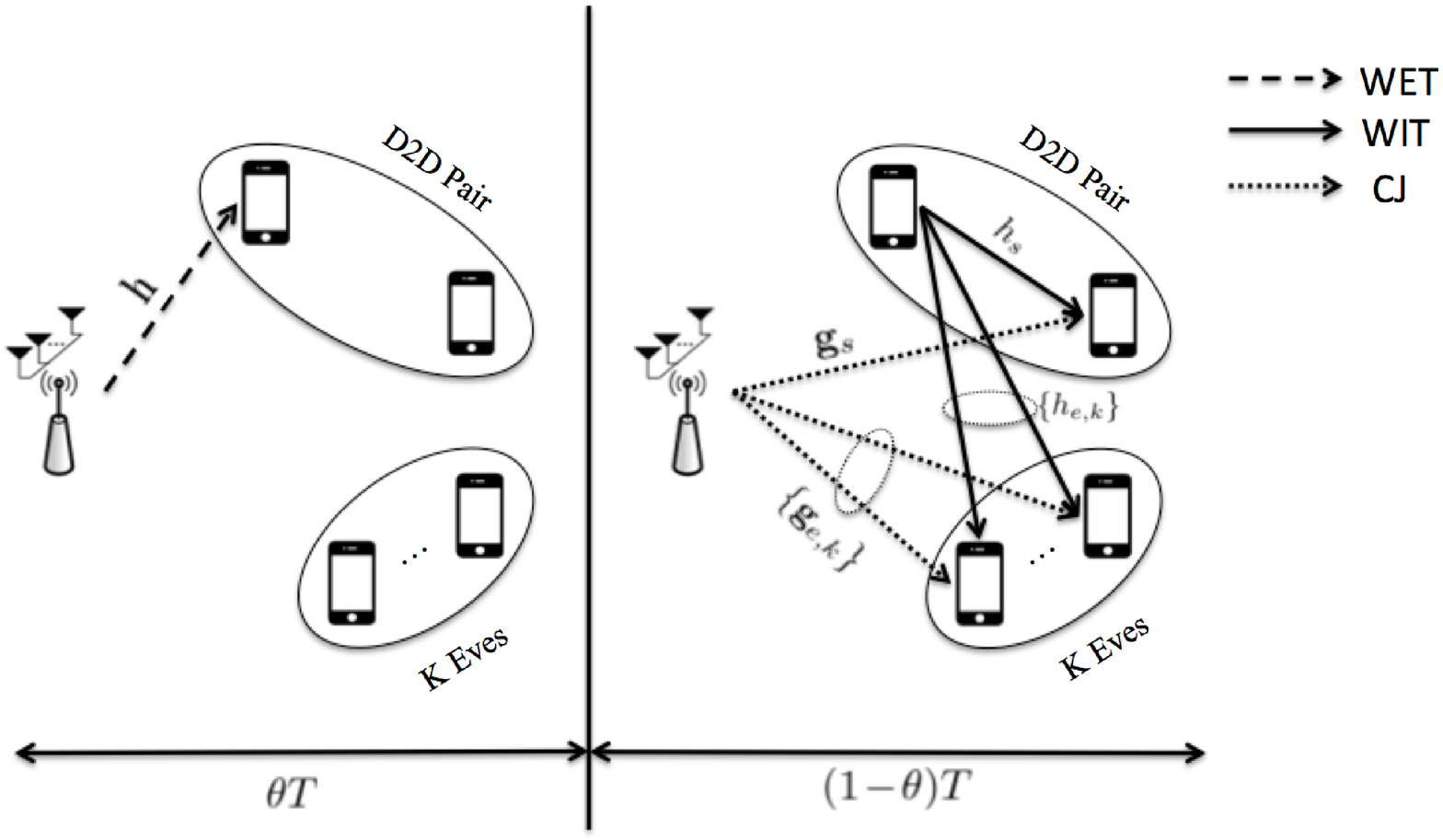}
	\caption{System model.}
	\label{fig:System_model}
\end{figure}

In this section, a secure wireless-powered D2D communication, shown in Fig. \ref{fig:System_model}, is considered. The communication scenario considered consists of a cellular hybrid BS, a D2D pair,\footnote{The D2D pair utilizes the same frequency of the cellular network in underlay mode. We ignore the D2D interference to the cellular network by assuming the D2D interference to be lower than the interference tolerance of the cellular network.} and multiple eavesdroppers. A secure D2D communication link is established between the D2D pair in the presence of $ K $ multiple eavesdroppers. In addition, a multi-antenna hybrid BS in the cellular network is employed to provide power for the D2D transmitter to guarantee the secure communications, and also to introduce jamming signals to interfere with the eavesdroppers. We assume that the D2D transmitter has no embedded power supply available and needs to be powered from the RF signals transmitted by the hybrid BS, which serves as a stable and reliable energy provider. We assume that the BS is powered constantly by a national grid or a micro grid. Moreover, the ``\emph{harvest-then-transmit}'' protocol is considered during a time period $ T $, shown in Fig. \ref{fig:System_model}, which is divided into two parts: (a) \emph{Wireless energy transfer (WET)}, i.e., in the time period of $ \theta T $ ($ 0 < \theta < 1 $), the BS provides power for the D2D transmitter by transmitting the RF signals; (b) \emph{Wireless information transfer and cooperative jamming (WITCJ)}, i.e., in the $ (1-\theta)T $ time period, the D2D transmitter employs the harvested energy to send information signals to the D2D receiver, and meanwhile, the BS transmits jamming signal to interfere with the eavesdroppers to enhance the secrecy rate. Without any loss of generality, we assume that $ T = 1 $ in this paper. It is assumed that the D2D pair and all eavesdroppers are equipped with a single antenna, while the hybrid BS is equipped with $ N_{T} $ transmit antennas. Let $ h_{s} $ and $ h_{e,k} $ denote the channel coefficients between the D2D transmitter and the D2D receiver as well as the $ k $-th eavesdropper, respectively. Furthermore, let $ \mathbf{h} \in \mathbb{C}^{1\times N_{T}} $, $ \mathbf{g}_{s} \in \mathbb{C}^{1\times N_{T}} $ and $ \mathbf{g}_{e,k} \in \mathbb{C}^{1\times N_{T}} $ be the channel coefficients between the BS and the D2D transmitter, the D2D receiver, as well as the $ k $-th eavesdropper, respectively.
During the WET phase, the BS employs the energy beamforming $ \mathbf{w} $ along with the direction of $ \mathbf{h} $, which can be written as $ \mathbf{w} = \frac{\mathbf{h}}{\|\mathbf{h}\|} $. Thus, the harvested energy at the D2D transmitter is given by
\begin{align}
E_{s} = \xi \theta P_{BS} \|\mathbf{h}\|^{2},
\end{align}
where $ \xi \in (0,1) $ denotes the energy harvesting efficiency of the D2D transmitter, and $ P_{BS} $ is the transmit power at the BS. Thus, the maximum transmit power available for the D2D transmitter during the $ (1-\theta)T $ time period can be written as
\begin{align}
p_{s}^{\textrm{max}} = \frac{E_{s}}{1-\theta} = \frac{\xi \theta P_{BS} \|\mathbf{h}\|^{2}}{1-\theta}.
\end{align}
During the WITCJ phase, while the secure communication is established between the D2D pair, the jamming signal is introduced by the BS to interfere the eavesdroppers. In order to confuse any potential eavesdroppers except the D2D receiver, the jamming signal is generated in the null space of the channel coefficients between the BS and the D2D user $ \mathbf{g}_{s} $, i.e., $ \mathbf{g}_{s}\mathbf{z} = \mathbf{0} $, where $ \mathbf{z} = \mathbf{T}\mathbf{v} $ is the jamming beamforming vector, $ \mathbf{T} \in \mathbb{C}^{N_{T} \times (N_{T} - 1)} $ is the orthogonal basis of the null space of $ \mathbf{g}_{s} $, and $ \mathbf{v} $ is a $ (N_{T} - 1) $ jamming signal vector with entries being independent and identically distributed (i.i.d) $ \mathcal{CN}(0,1) $ variables. In this paper, we assume that the jamming power (also equals to $ P_{BS} $) is uniformly distributed among the $ N_{T} - 1 $ dimensions.
Based on the above analysis and assumption, the instantaneous channel capacity at the D2D receiver and the $ k $-th eavesdropper can be, respectively, written as
\begin{align}
C_{s} = \log \bigg(1+ \frac{p_{s}|h_{s}|^{2}}{\sigma_{s}^{2}}\bigg),
\end{align}
and
\begin{align}
C_{e,k} = \log \bigg( 1+ \frac{p_{s}|h_{e,k}|^{2}}{\frac{P_{BS}}{N_{T} - 1} \|\mathbf{g}_{e,k}\mathbf{T}\|^{2}+\sigma_{e}^{2}} \bigg),
\end{align}
where $ p_{s} $ is the transmit power at the D2D transmitter, $ \sigma_{s}^{2} = \sigma_{I}^{2} + \sigma_{n}^{2} $, $ \sigma_{I}^{2} $ is the interference power from cellular network,\footnote{It is assumed that the interference from cellular network is an ergodic process, i.e., the statistical average can be estimated by averaging over time, and D2D user can estimate the interference power $ \sigma_{I}^{2} $. For the detailed discussion on interference modeling, interested reader are referred to \cite{Ren}.} and $ \sigma_{n}^{2} $ and $ \sigma_{e}^{2} $ are the noise variances of additive white Gaussian noise (AWGN) at the D2D receiver and the $ k $-th eavesdropper, respectively.
Without loss of generality, the worst case scenario is considered in this paper, where it is assumed that $ \sigma_{e}^{2} = 0 $.
Thus, the secrecy capacity can be written as
\begin{align}
C = [C_{s} - \max_{k} C_{e,k}]^{+}.
\end{align}
In addition, the secrecy rate $ R $ in this paper is defined as the difference between communication rate of main channel $ R_{s} $ ($ R_{s} \leq C_{s} $) and the maximum eavesdropping rate, i.e., $ \max_{k \in [1,K]} R_{e,k} $, which is given by
\begin{align}
R_{e} = \max_{k \in [1,K]} R_{e,k} = \log(1+\rho_{e}),
\end{align}
where $ \rho_{e} \geq 0 $ denotes the associated signal-to-noise ratio (SNR) threshold for the eavesdroppers.\footnote{The associated SNR threshold $ \rho_{e} $ is an optimization variable which is introduced to control the information leakage levels at the eavesdroppers. The D2D transmitter optimizes this variable to maximize its utility function.} Thus, the secrecy rate for the D2D network can be expressed as
\begin{align}
R = R_{s} - R_{e} = R_{s} - \log(1+\rho_{e}).
\end{align}
It is noted that $ R_{s} $ should satisfy $ R_{s} \leq C_{s} $ to guarantee reliable transmissions, which implies that $ R_{s} $ is upper-bounded by $ C_{s} $.
In order to maximize $ R $, we consider $ R_{s} = C_{s} $, and thus
\begin{align}
R = C_{s} -  \log(1+\rho_{e}).
\end{align}
Due to the randomness of the eavesdroppers' channels,  it is difficult to estimate the accuracy CSI of the eavesdroppers. In this case, we assume that the D2D transmitter only knows statistical CSI of the eavesdroppers. Under this assumption, the secrecy outage probability is considered as a secrecy metric, which is defined as follows:
\begin{align}\label{eq:SOP_ori}
p_{out} = \textrm{Pr} \{ C_{s} - \max_{k} C_{e,k} < R \}.
\end{align}
We assume that $ h_{e,k} $ is i.i.d. complex Gaussian variable with variance $ \gamma_{e}^{2} $, and the entries of $ \mathbf{g}_{e,k}\mathbf{T} $ are i.i.d. complex Gaussian variables with variance $ \delta_{e}^{2} $. Thus, the secrecy outage probability \eqref{eq:SOP_ori} can be derived as
\begin{align}\label{eq:Secrecy_outage_probability_solution}
p_{out} &~= \textrm{Pr} \{ C_{s} - \max_{k} \{ C_{e,k} \} < R \} \nonumber\\
&~= \textrm{Pr} \{ \max_{k} \{ C_{e,k} \} > \log(1+\rho_{e}) \}  \nonumber\\
&~ = 1 - \textrm{Pr} \left\{ \max_{k} \bigg\{ \frac{p_{s}(N_{T} - 1)|h_{e,k}|^{2}}{P_{BS}\|\mathbf{g}_{e,k}\mathbf{T}\|^{2}} \bigg\} \leq \rho_{e} \right\} \nonumber\\
&~ = 1 -  \bigg[ \textrm{Pr} \bigg(  \frac{p_{s}(N_{T} - 1)|h_{e,k}|^{2}}{P_{BS}\|\mathbf{g}_{e,k}\mathbf{T}\|^{2}}  \leq \rho_{e} \bigg) \bigg]^{K}  \nonumber\\
&~ = 1 \!-\! \bigg\{ 1 \!-\! \bigg[ 1\!+\! \rho_{e}\bigg( \frac{P_{BS} \gamma_{e}^{2}}{p_{s}\delta_{e}^{2}(N_{T}\!-\!1)} \bigg) \bigg]^{1 \!-\! N_{T}} \bigg\}^{K}.
\end{align}
From \eqref{eq:Secrecy_outage_probability_solution}, we can observe that $ p_{out} $ decreases and asymptotically reaches zero when $ N_{T} \rightarrow \infty $.

Based on the energy transfer time allocation $ \theta $ and secrecy rate $ R $, the secrecy throughput is defined as the average number of confidential messages received at the D2D receiver per unit time, and can be given as
\begin{align}
\!\!\!\! T_{s} \!=\! (1\!-\!\theta) R \!=\! (1\!-\!\theta)\bigg[\!\log \bigg(1\!+\!\frac{p_{s}|h_{s}|^{2}}{\sigma_{s}^{2}}\bigg) \!-\! \log (1\!+\!\rho_{e})\!\bigg].
\end{align}
\section{Game Theoretical Based Secure Wireless Powered D2D Communications}\label{section Game_theoretical_scheme}
In practice, both WET and WITCJ processes can belong to different service providers and act strategically. Thus, the D2D transmitter has to pay a price as incentive to the BS for WET service to facilitate its secure communications. In order to efficiently exploit the WET interaction between the BS and the D2D transmitter, we model both WET and WITCJ processes as a \emph{Stackelberg} game, where two different schemes are considered as \textit{energy trading} and \textit{non-energy trading}. When the two WET and WITCJ processes belong to the same service provider, we consider the third scheme called \textit{social wellfare} optimization.
\subsection{Energy Trading based Stackelberg Game}\label{section Energy_trading_stackelberg_game}
In this subsection, we consider an energy trading framework for the secure wireless-powered D2D CJ-aided network using \emph{Stackelberg} game. Particularly, the strategic interactions between the D2D network and the cellular network are taken into consideration. In the formulated game, the D2D transmitter plays a leader role who purchases the energy service from the hybrid BS in cellular network by offering a price to guarantee WET and secure WITCJ processes during time period $ T $.
The D2D transmitter optimizes its energy price as well as energy transfer time allocation to maximize its utility function.
On the other hand, the hybrid BS is modelled as the follower of this game, which decides their optimal transmit powers based on the energy price paid by the D2D transmitter to maximize its own utility function.
Letting $ \lambda $ denote the energy price released by the D2D transmitter, mathematically, the payment to be paid by the D2D transmitter to the BS can be expressed as
\begin{align}
E(\theta,\lambda,P_{BS}) = \lambda \theta P_{BS} \|\mathbf{h}\|^{2}.
\end{align}
Then, the utility function of the D2D transmitter can be defined as the difference between the benefits of the achievable secrecy throughput and its payment to the BS, which is given by
\begin{align}
U_{L}(\theta,\lambda,P_{BS}) = \mu T_{s} - E,
\end{align}
where $ \mu > 0 $ is the gain per unit throughput for the D2D transmitter. Thus, the optimization problem the D2D network utility (leader-level game) is formulated as: \\
\textbf{Leader Level}
\begin{align}\label{eq:Energy_trading_leader_level}
\max_{p_{s},\lambda,\rho_{e},\theta}&~ U_{L}(\theta,\lambda,P_{BS}), ~s.t. ~\rho_{e} \geq 0, ~\lambda \geq 0,~p_{out} \leq \varepsilon,\nonumber\\&~ 0 \leq p_{s} \leq p_{s}^{\max}, ~ 0 < \theta < 1.
\end{align}
Note that the optimal value of $ \theta $ can be neither 0 nor 1 as both values result in zero secrecy throughput. Moreover, the hybrid BS in the cellular network is modeled as the follower who maximizes its own profit, which can be expressed as
\begin{align}
U_{BS}(\theta, \lambda, P_{BS}) = \theta (\lambda  P_{BS} \|\mathbf{h}\|^{2} - \mathcal{F}(P_{BS})),
\end{align}
where $ \mathcal{F}(P_{BS}) $ is used to model the cost of the BS per unit time for wirelessly charging. In this paper, we consider the following quadratic model\footnote{Note that the quadratic function shown in \eqref{eq:Quadratic_function_power_market} has been applied in the energy market to model the energy cost \cite{Schober_TSG2010_DSM_GT}.} for the cost function of the PBs.
\begin{align}\label{eq:Quadratic_function_power_market}
\mathcal{F}(x) = A x^{2} + B x
\end{align}
where $ A > 0 $ and $ B > 0 $ are constants.  \\
Thus, the optimization problem for the hybrid BS or the follower-level game is given by \\
\textbf{Follower Level}
\begin{align}\label{eq:Energy_trading_follower_game}
\max_{P_{BS}} U_{BS} &= \lambda \theta P_{BS} \|\mathbf{h}\|^{2} - \theta (AP_{BS}^{2} + BP_{BS}), \nonumber\\ s.t.&~ P_{BS} \geq 0.
\end{align}
The \emph{Stackelberg} game for the considered WPCN-aided secure D2D network has been formulated by combining problems \eqref{eq:Energy_trading_leader_level} and \eqref{eq:Energy_trading_follower_game}.\footnote{Generally, the BS provides two services for the D2D pair: energy service and CJ service. In our formulated game, the BS can release these two services by one combined price. After receiving payment for the energy service, the interference service will be included in the package offered by the BS. So the interference service is used as incentive to attract energy buyers.} The D2D transmitter (the leader) aims to solve the problem \eqref{eq:Energy_trading_leader_level}, whereas the BS (the follower) aims to solve the problem \eqref{eq:Energy_trading_follower_game}.
For this game, the subsequent part is to find the \emph{Stackelberg} equilibrium, which can be formally defined as:
\begin{definition}\label{definition Stackelberg_equilibrium}
	Let $ (\theta^{\textrm{opt}}, \lambda^{\textrm{opt}}) $ denote the solutions to the problem \eqref{eq:Energy_trading_leader_level}, while $ P_{BS}^{\textrm{opt}} $ represents the solution to the problem \eqref{eq:Energy_trading_follower_game}, then, the tuple ($ \theta^{\textrm{opt}}, \lambda^{\textrm{opt}},P_{BS}^{\textrm{opt}} $) is the \emph{Stackelberg} equilibrium of the formulated game provided that the following conditions are satisfied
	\begin{align}
	&~U_{L}(\theta^{\textrm{opt}}, \lambda^{\textrm{opt}},P_{BS}^{\textrm{opt}}) \geq U_{L}(\theta, \lambda,P_{BS}^{\textrm{opt}}), \\
	&~U_{BS}(\theta^{\textrm{opt}}, \lambda^{\textrm{opt}},P_{BS}^{\textrm{opt}}) \geq U_{BS}(\theta^{\textrm{opt}}, \lambda^{\textrm{opt}},P_{BS}),
	\end{align}
	for $ 0<\theta <1 $, $ \lambda \geq 0 $, and $ P_{BS} \geq 0 $.
\end{definition}

Exploiting the \emph{Stackelberg} equilibrium definition in \emph{Definition} \ref{definition Stackelberg_equilibrium}, we will derive the Stackelberg equilibrium of the formulated game by analyzing the optimal strategies for the D2D transmitter and the hybrid BS to maximize their own utility functions.
It can be seen from \eqref{eq:Energy_trading_follower_game} that for a given energy price $ \lambda $ and the energy transfer time allocation $ \theta $, the utility function of the BS is a quadratic function with respect to $ P_{BS} $ and the constraint is affine, which indicate that \eqref{eq:Energy_trading_follower_game} is a convex problem. Thus, the following \emph{lemma} is required to obtain its optimal solution:
\begin{lemma}\label{lemma Optimal_P_BS}
	For given $ \lambda $ and $ \theta $,  the optimal solution to \eqref{eq:Energy_trading_follower_game} can be achieved as
	\begin{align}\label{eq:Optimal_P_BS_energy_trading}
	P_{BS}^{\textrm{opt}} = \left[\frac{\lambda \|\mathbf{h}\|^{2} - B}{2 A}\right]^{+}.
	\end{align}
\end{lemma}
\begin{IEEEproof}
		It can be verified that the problem \eqref{eq:Energy_trading_follower_game} is convex in terms of $ P_{BS} $. Thus, we take into account the first derivative to the objective function in \eqref{eq:Energy_trading_follower_game} equals to zero as
	\begin{align}
	\frac{\partial U_{BS}}{\partial P_{BS}} & = \lambda \theta \|\mathbf{h}\|^{2} - 2 \theta A P_{BS} - \theta B = 0, \nonumber\\
	&~~~~ \Rightarrow P_{BS} = \frac{\lambda \|\mathbf{h}\|^{2} - B}{2 A}.
	\end{align}
	Combining with the condition $ P_{BS} > 0 $, we have completed the proof of \emph{Lemma} \ref{lemma Optimal_P_BS}.
\end{IEEEproof}
Here, we first consider the optimal solutions to the SNR threshold $ \rho_{e} $ and the D2D transmit power $ p_{s} $ by the following problem for the given $ \lambda $, $ \theta $ and $ P_{BS} $,
\begin{align}\label{eq:Energy_trading_leader_game_reformulated}
\max_{p_{s},\rho_{e}}  \frac{1\!+\!\frac{p_{s}|h_{s}|^{2}}{\sigma_{s}^{2}}}{1\!+\!\rho_{e}} ~ 
s.t. ~ 0 \! \leq \! p_{s} \! \leq \! p_{s}^{\max},~ \rho_{e} \! \geq \! 0,~ p_{out} \! \leq \! \varepsilon.
\end{align}
In order to obtain the optimal solutions to \eqref{eq:Energy_trading_leader_game_reformulated}, the following \emph{theorem} is introduced:
\begin{theorem}\label{theorem Optmal_p_s_and_rho_e}
	For given $ \theta $ and $ P_{BS} $, the optimal transmit power at the D2D transmitter $ p_{s}^{\textrm{opt}} $ and the eavesdroppers' SNR threshold $ \rho_{e}^{\textrm{opt}} $ are expressed in terms of the closed-form solution, respectively, as
	\begin{align}
	p_{s}^{\textrm{opt}} = \frac{\theta \xi P_{BS} \|\mathbf{h}\|^{2}}{1-\theta},
	\end{align}
	and
	\begin{align}\label{eq:Optimal_rho_e}
	\rho_{e}^{\textrm{opt}} = \frac{\theta \xi \| \mathbf{h} \|^{2} (N_{T} - 1) \{ [1 - (1-\varepsilon)^{\frac{1}{K}} ]^{\frac{1}{1-N_{T}}} - 1 \} \delta_{e}^{2} }{(1-\theta)\gamma_{e}^{2}}.
	\end{align}
\end{theorem}
\begin{IEEEproof}
First, we fix $ p_{s} $ to obtain the solution to $ \rho_{e} $ from \eqref{eq:Secrecy_outage_probability_solution}, which is the function with respect to $ p_{s} $, as follows:
\begin{align}\label{eq:Solution_rho_e_for_p_s}
\rho_{e}(p_{s}) = \frac{(N_{T} - 1)\{ [1 - (1-\varepsilon)^{\frac{1}{K}} ]^{\frac{1}{1-N_{T}}} - 1 \}\delta_{e}^{2}}{P_{BS}\gamma_{e}^{2}}p_{s}.
\end{align}	
It is noted from \eqref{eq:Energy_trading_leader_game_reformulated} that $ |h_{s}|^{2} > \frac{(N_{T} - 1)\{ [1 - (1-\varepsilon)^{\frac{1}{K}} ]^{\frac{1}{1-N_{T}}} - 1 \}\delta_{e}^{2}}{P_{BS}\gamma_{e}^{2}} $ for positive achievable secrecy rate. The objective function in \eqref{eq:Energy_trading_leader_game_reformulated} is monotonically increasing with respect to $ p_{s} $. Thus, the optimal $ p_{s}^{\textrm{opt}} $ for any given energy transfer time allocation $ \theta $ can be expressed as
\begin{align}\label{eq:Optimal_p_s}
p_{s}^{\textrm{opt}} = \frac{\theta \xi P_{BS} \|\mathbf{h}\|^{2}}{1-\theta}.
\end{align}
Replacing \eqref{eq:Solution_rho_e_for_p_s} with \eqref{eq:Optimal_p_s}, we have the optimal solution $ \rho_{e}^{\textrm{opt}} $ in \eqref{eq:Optimal_rho_e}.
\end{IEEEproof}
Next, we solve the problem \eqref{eq:Energy_trading_leader_level} by substituting $ P_{BS} $ with $ P_{BS}^{\textrm{opt}} $ given in \eqref{eq:Optimal_P_BS_energy_trading}.
Based on \emph{Lemma} \ref{lemma Optimal_P_BS} and \emph{Theorem} \ref{theorem Optmal_p_s_and_rho_e}, we rewrite the optimization problem \eqref{eq:Energy_trading_leader_level} with $ p_{s}^{\textrm{opt}} $, $ P_{BS}^{\textrm{opt}} $ and $ \rho_{e}^{\textrm{opt}} $ as follows:
\begin{align}\label{eq:Follower_game_energy_trading_with_optimal}
\max_{\lambda, \theta}& ~ U_{L}(\theta,\lambda) = \mu(1-\theta) \log\bigg( \frac{1+\frac{\theta \xi P_{BS}^{\textrm{opt} }\|\mathbf{h}\|^{2}|h_{s}|^{2}}{(1-\theta)\sigma_{s}^{2}}}{1+\rho_{e}^{\textrm{opt}}} \bigg) \nonumber\\&~~~~~~~~~~- \lambda \theta P_{BS}^{\textrm{opt}} \|\mathbf{h}\|^{2} \nonumber\\
&~ = \mu(1-\theta) \bigg[ \log\bigg( 1 + \frac{\theta \xi |h_{s}|^{2} }{(1-\theta)\sigma_{s}^{2}} \frac{\lambda \|\mathbf{h}\|^{2} - B}{2 A} \|\mathbf{h}\|^{2}  \bigg) \nonumber\\ &~~~~~~~~  - \log (1+\rho_{e}^{\textrm{opt}}) \bigg] - \lambda \theta \bigg(\frac{\lambda \|\mathbf{h}\|^{2} - B}{2 A}\bigg) \|\mathbf{h}\|^{2}, \nonumber\\
&~ = \mu(1\!-\!\theta) \bigg[ \log\bigg( 1 \!+\! \frac{ \xi |h_{s}|^{2}}{(1-\theta)\sigma_{s}^{2}} \bigg( \frac{\lambda\theta \|\mathbf{h}\|^{4}}{2 A} \!-\! \frac{\theta B \|\mathbf{h}\|^{2}}{2 A} \bigg) \bigg) \nonumber\\ &~~~~~~~ - \log(1+\rho_{e}^{\textrm{opt}}) \bigg] - \frac{\lambda^{2} \theta \|\mathbf{h}\|^{4}}{2 A} + \frac{\lambda \theta B \|\mathbf{h}\|^{2}}{2 A}, \nonumber\\
s.t. &~ \lambda \geq 0,~ 0 \leq \theta \leq 1.
\end{align}
From \eqref{eq:Follower_game_energy_trading_with_optimal}, it is hard to find the optimal expressions for $ \theta $ and $ \lambda $ at the same time due to the complexity of its objective function. In order to tackle this issue, we optimally solve \eqref{eq:Follower_game_energy_trading_with_optimal} in two steps.
We first find the optimal closed-form solution to $ \lambda $ for a given $ \theta $. Then, the optimal value for $ \theta $ can be obtained via one-dimension line search. Now we rewrite \eqref{eq:Follower_game_energy_trading_with_optimal} for a given $ \theta $ as
\begin{align}\label{eq:Energy_trading_U_L_with_lambda}
\max_{\lambda} &~U_{L}(\lambda) = a \bigg[ \log (1 + d (\lambda C - 2 D) ) - \log(1+\rho_{e}^{\textrm{opt}}) \bigg] \nonumber\\ &~~~~~~ - \lambda^{2} C + 2 \lambda D, \nonumber\\
s.t. &~ \lambda \geq 0,
\end{align}
where
\begin{align}
C = \frac{\theta \|\mathbf{h}\|^{4}}{2 A}, ~ D = \frac{\theta B \|\mathbf{h}\|^{2}}{4 A},~ a = \mu (1-\theta),~ d = \frac{\xi |h_{s}|^{2}}{(1-\theta) \sigma_{s}^{2}}. \nonumber
\end{align}
We solve problem \eqref{eq:Energy_trading_U_L_with_lambda} by introducing the following \emph{theorem}:
\begin{theorem}
	The optimal solution to problem \eqref{eq:Energy_trading_U_L_with_lambda} can be derived as
	\begin{align}\label{eq:Optimal_lambda_energy_trading}
\lambda^{\textrm{opt}} = \frac{-(1 - 3 d D) + \sqrt{(1 - d D )^{2} + 2 a d^{2} C}}{2 d C}.
	\end{align}
\end{theorem}
\begin{IEEEproof}
It is easily verified that the objective function in \eqref{eq:Energy_trading_U_L_with_lambda} is a concave function, thus, in order to find the optimal solution to $ \lambda $, we consider its first-order derivatives that equals to zero,
\begin{align}
\frac{\partial U_{L}}{\partial \lambda} = \frac{a d C}{1+ d (\lambda C - D) - d D} - 2 \lambda C + 2 D = 0.
\end{align}
After some mathematical simplifications, we have
\begin{align}
2 d (\lambda C - D)^{2} + 2 (1 - d D)(\lambda C - D) - a d C = 0.
\end{align}
Let $ x = \lambda C - D $, we get
\begin{align}
2 d x^{2} + 2 (1 - d D) x - a d C = 0.
\end{align}
The optimal solution to $ x $ is easily achieved by solving the above equation. Thus, the associated optimal solutions to $ \lambda $ can be derived as
\begin{align}\label{eq:Both_solutions_lambda}
\begin{cases}
\lambda_{1} = \frac{-(1 - 3 d D) + \sqrt{(1 - d D )^{2} + 2 a d^{2} C}}{2 d C}, \\
\lambda_{2} = \frac{-(1 - 3 d D) - \sqrt{(1 - d D )^{2} + 2 a d^{2} C}}{2 d C}.
\end{cases}
\end{align}
Now, let us verify the validity of both solutions shown in \eqref{eq:Both_solutions_lambda}. The objective function \eqref{eq:Energy_trading_U_L_with_lambda} includes the logarithm term, which should be non-negative. Thus, we check the validity of these solutions by substituting $ \lambda_{1} $ and $ \lambda_{2} $ into the logarithm term of \eqref{eq:Energy_trading_U_L_with_lambda}, respectively. We first check $ \lambda_{1} $ as follows:
\begin{align}
& 1 + d\bigg(\frac{-(1 - 3 d D) + \sqrt{(1 - d D )^{2} + 2 a d^{2} C}}{2 d} - 2 D \bigg) \nonumber\\
& = 1 + d\bigg(\frac{-(1 -  d D) + \sqrt{(1 - d D )^{2} + 2 a d^{2} C}}{2 d} \bigg) \nonumber\\
& > 1 + d\bigg(\frac{-(1 -  d D) + | 1 - d D | }{2 d} \bigg) \geq 1.
\end{align}
Similarly, we check $ \lambda_{2} $ as
\begin{align}
& 1 + d\bigg(\frac{-(1 - 3 d D) - \sqrt{(1 - d D )^{2} + 2 a d^{2} C}}{2 d} - 2 D \bigg) \nonumber\\
& = 1 + d\bigg(\frac{-(1 -  d D) - \sqrt{(1 - d D )^{2} + 2 a d^{2} C}}{2 d} \bigg) \nonumber\\
& < 1 + d\bigg(\frac{-(1 -  d D) - | 1 - d D |}{2 d} \bigg) \leq 1.
\end{align}
Thus, $ \lambda_{1} $ is a valid stationary point. Due to the concavity of the objective function in \eqref{eq:Energy_trading_U_L_with_lambda}, its second-order derivatives with respective to $ \lambda $ is less than zero, which indicates that its maximum value is the stationary point $ \lambda_{1} $. Also, it is easily verified that $ \lambda_{1} > 0 $, which satisfied the constraint \eqref{eq:Energy_trading_U_L_with_lambda}. Thus, the optimal solution to the problem \eqref{eq:Energy_trading_U_L_with_lambda}, denoted by $ \lambda^{\textrm{opt}} $ is the stationary point $ \lambda_{1} $, which completes the proof.
\end{IEEEproof}

We have already achieved the optimal energy price $ \lambda^{\textrm{opt}} $ of the D2D transmitter for a given energy transfer time allocation $ \theta $.
Substituting $ \lambda^{\textrm{opt}} $ given in \eqref{eq:Optimal_lambda_energy_trading} into the problem \eqref{eq:Follower_game_energy_trading_with_optimal}, we have the following problem regarding $ \theta $:
\begin{align}\label{eq:U_L_with_theta}
\max_{\theta} &~ U_{L}(\theta,\lambda^{\textrm{opt}}) = \mu (1-\theta) \bigg[ \log\bigg(1+ t_{1}\frac{\theta}{1-\theta}\bigg) \nonumber\\&~~~~~~~~~~~~~~~~~~ - \log\bigg(1+t_{2} \frac{\theta}{1-\theta}\bigg) \bigg] - \theta t_{3} ,\nonumber\\
s.t. &~ 0<\theta<1,
\end{align}
where
\begin{align}
t_{1} & = \frac{\xi |h_{s}|^{2} (\lambda^{\textrm{opt}}\|\mathbf{h}\|^{4} - B \|\mathbf{h}\|^{2})}{2 A \sigma_{s}^{2}},\nonumber\\ 
t_{2} & = \frac{ \xi \| \mathbf{h} \|^{2} (N_{T} - 1) \{ [1 - (1-\varepsilon)^{\frac{1}{K}} ]^{\frac{1}{1-N_{T}}} - 1 \} \delta_{e}^{2} }{\gamma_{e}^{2}},\nonumber\\
t_{3} &= \frac{\lambda^{2} \|\mathbf{h}\|^{4}}{2 A} - \frac{\lambda B \|\mathbf{h}\|^{2}}{2 A}. \nonumber
\end{align}
From \eqref{eq:U_L_with_theta}, it can be seen that $ U_{L} $ is a concave function with respect to $ \theta $ when $ t_{1} > t_{2} $ holds to guarantee the positive achievable secrecy rate. Thus, \eqref{eq:U_L_with_theta} is a convex problem. In order to obtain the optimal solution to $ \theta $, we consider the following equation by taking the first-order derivative of $ U_{L} $ with respect to $ \theta $ as
\begin{align}\label{eq:First_order_derivative_U_L_with_theta}
&  \mu \log\bigg( \frac{(t_{1} - 1)\theta + 1}{(t_{2} - 1)\theta + 1} \bigg) \nonumber\\ 
 &~~~= \mu (1-\theta) \bigg[ \frac{t_{1} - 1}{(t_{1} - 1)\theta + 1} - \frac{t_{2} - 1}{(t_{2} - 1)\theta + 1}\bigg] - t_{3}.
\end{align}
We set the left-hand-side (LHS) and right-hand-side (RHS) of \eqref{eq:First_order_derivative_U_L_with_theta} as $ f(\theta) $ and $ g(\theta) $, respectively. It is easily shown that $ f(\theta) $ is monotonically increasing function with $ \theta $, whereas $ g(\theta) $ is monotonically decreasing function with $ \theta $. Thus, the optimal solution of $ \theta $ can be achieved by iteratively updating $ \theta $ from the initial value $0$ to the value such that $ f(\theta) = g(\theta) $, which is of lower complexity than one-dimensional (1D) line search.
With $ f(0) = 0 $, $ f(1) = \mu\log\left(\frac{t_{1}}{t_{2}}\right) > 0 $, $ g(0) = \mu (t_{1} - t_{2}) - t_{3} > 0 $, and $ g(1) = -t_{3} < 0 $, this optimal solution $ \theta^{\textrm{opt}} > 0 $, located at the intersection, is feasible, and also satisfies the constraint in \eqref{eq:U_L_with_theta}. Thus, we denote the optimal solution to \eqref{eq:U_L_with_theta} as follows:
\begin{align}\label{eq:Optimal_theta_energy_trading}
\theta^{\textrm{opt}} = \arg \max_{\theta \in (0,1)} U_{L}(\theta,\lambda^{\textrm{opt}}).
\end{align}
With this we complete the derivation of the \emph{Stackelberg} game's  equilibrium $(P_{BS}^{\textrm{opt}},\lambda^{\textrm{opt}},\theta^{\textrm{opt}}),$ which are presented in \eqref{eq:Optimal_P_BS_energy_trading}, \eqref{eq:Optimal_lambda_energy_trading}, and \eqref{eq:Optimal_theta_energy_trading}, respectively.


\subsection{Non-Energy Trading based Stackelberg Game}
In Section \ref{section Energy_trading_stackelberg_game}, we formulate the energy trading between the hybrid BS and the D2D transmitter. 
As a comparison, in this subsection, we propose non-energy trading based \emph{Stackelberg} game formulation to exploit the interaction between the BS and the D2D transmitter with a fixed energy transfer time allocation $ \theta $. In this game, the hybrid BS is modelled as the leader who determines the energy price to maximize its utility which is defined as the difference between the total D2D transmitter payment and the quadratic energy cost shown in \eqref{eq:Quadratic_function_power_market}.
Thus, the leader level of this \emph{Stackelberg} game is written as\footnote{This leader level game is similar to the follower level of the energy trading based \emph{Stackelberg} game \eqref{eq:Energy_trading_follower_game}. The only difference between these two games is that this leader game is to decide the energy price $ \lambda $ to maximize its utility, whereas the follower game \eqref{eq:Energy_trading_follower_game} determines the optimal transmit power $ P_{BS} $ to maximize its utility.} \\
\textbf{Leader Level}
\begin{align}\label{eq:No_energy_trading_leader}
\max_{\lambda}&~ U_{BS} = \lambda \theta P_{BS} \|\mathbf{h}\|^{2} - \theta (AP_{BS}^{2} + BP_{BS}),\nonumber\\
s.t. &~\lambda \geq 0.
\end{align}
In addition, the D2D transmitter plays the follower's role to guarantee the secure communication by using the power released by the BS, in which it aims to maximize the difference between the benefit of the achievable secrecy throughput and its payment to the BS for wireless power transfer.
Thus, the follower level of this \emph{Stackelberg} game is given by \\
\textbf{Follower Level}
\begin{align}\label{eq:No_energy_trading_follower}
\max_{P_{BS},p_{s},\rho_{e}} &~ U_{L}(\theta,P_{BS},p_{s},\rho_{e}), \nonumber \\ s.t. &~ \rho_{e} \geq 0,~p_{out} \leq \varepsilon,~ 0 \leq p_{s}\leq p_{s}^{\max},
\end{align}
where
\begin{align}\label{eq:No_energy_trading_utility_function}
& U_{L}(\theta,P_{BS},p_{s},\rho_{e}) = \mu T_{s} - E \nonumber\\
&~~= \mu (1-\theta) \bigg[ \log \bigg( 1 + \frac{p_{s}|h_{s}|^{2}}{\sigma_{s}^{2}} \bigg) - \log(1+\rho_{e}) \bigg] \nonumber\\ &~~~~~~~~~~ - \lambda \theta P_{BS} \|\mathbf{h}\|^{2}.
\end{align}
We first employ the same method to obtain the optimal solutions $ p_{s}^{\textrm{opt}} $, $ \rho_{e}^{\textrm{opt}} $ and $ \theta^{\textrm{opt}} $  shown in Section \ref{section Energy_trading_stackelberg_game}. Then, we focus on the optimal solution of the BS transmit power by solving problem \eqref{eq:No_energy_trading_follower}. It can be verified that the utility function in \eqref{eq:No_energy_trading_utility_function} is a concave function with respect to $ P_{BS} $. Now, we set its first-order derivative equal to zero,
\begin{align}
\frac{\partial U_{L}}{\partial P_{BS}} = \frac{\mu (1 - \theta) \frac{\xi \theta \|\mathbf{h}\|^{2} |h_{s}|^{2}}{(1-\theta)\sigma_{s}^{2}} }{1 + \frac{\xi \theta \|\mathbf{h}\|^{2} |h_{s}|^{2}}{(1-\theta)\sigma_{s}^{2}}  P_{BS}} - \lambda \theta \|\mathbf{h}\|^{2} = 0.
\end{align}
After some mathematical derivations, the optimal power allocation of the BS with respect to $ \lambda $ is given by
\begin{align}\label{eq:Optimal_P_BS_no_energy_trading}
P_{BS}(\lambda) = \bigg[ \frac{\mu (1 - \theta)}{\lambda \theta \|\mathbf{h}\|^{2}} - \frac{(1 - \theta)\sigma_{s}^{2}}{\xi \theta \|\mathbf{h}\|^{2}|h_{s}|^{2}} \bigg]^{+}.
\end{align}
Substituting \eqref{eq:Optimal_P_BS_no_energy_trading} into \eqref{eq:No_energy_trading_leader} leads to the following:
\begin{align}\label{eq:Utility_no_energy_trading_with_optimal_P_{BS}}
U_{BS} &~ = \lambda \theta \bigg( \frac{\mu (1 - \theta)}{\lambda \theta \|\mathbf{h}\|^{2}} - \frac{(1 - \theta)\sigma_{s}^{2}}{\xi \theta \|\mathbf{h}\|^{2} |h_{s}|^{2}} \bigg) \|\mathbf{h}\|^{2} \nonumber\\ &~~~~~~~~ - \theta A \bigg( \frac{\mu (1 - \theta)}{\lambda \theta \|\mathbf{h}\|^{2}} - \frac{(1 - \theta)\sigma_{s}^{2}}{\xi \theta \|\mathbf{h}\|^{2} |h_{s}|^{2}} \bigg)^{2} \nonumber\\ &~~~~~~~~ - \theta B \bigg( \frac{\mu (1 - \theta)}{\lambda \theta \|\mathbf{h}\|^{2}} - \frac{(1 - \theta)\sigma_{s}^{2}}{\xi \theta \|\mathbf{h}\|^{2} |h_{s}|^{2}} \bigg) .
\end{align}
The utility function \eqref{eq:Utility_no_energy_trading_with_optimal_P_{BS}} is a concave function with respect to $ \lambda $. Taking the first-order derivative to \eqref{eq:Utility_no_energy_trading_with_optimal_P_{BS}} equal to zero, we get
\begin{align}\label{eq:Equation_for_lambda_no_energy_trading}
\frac{\partial U_{BS}}{\partial \lambda} &~ \!\!\!=\! -\theta Y \|\mathbf{h}\|^{2} \!-\! 2 \theta A \bigg(\!\frac{X}{\lambda} \!-\! Y \!\bigg) \bigg(\! -\! \frac{X}{\lambda^{2}} \bigg) \!-\! \theta B \bigg(\! -\! \frac{X}{\lambda^{2}} \!\bigg) \!=\! 0, \nonumber\\
&~ \Rightarrow
\lambda^{3} + \frac{2 A X Y - B X}{Y \|\mathbf{h}\|^{2}} \lambda - \frac{2 A X^{2}}{Y \|\mathbf{h}\|^{2}} = 0,
\end{align}
where
\begin{align}
X = \frac{\mu (1 - \theta)}{\theta \|\mathbf{h}\|^{2}},~ Y = \frac{(1 - \theta) \sigma_{s}^{2}}{\xi \theta \|\mathbf{h}\|^{2} |h_{s}|^{2}}
\end{align}
 It is observed that \eqref{eq:Equation_for_lambda_no_energy_trading} is a cubic equation, which can be solved in terms of closed-form solution of $ x $ by using Cardano's formula \cite{Xiaobin_Huang_CL_2014_FD},
\begin{align}\label{eq:Optimal_lambda}
\lambda^{\textrm{opt}} =  e^{j \angle \lambda_{1}} \sqrt[3]{|\lambda_{1}|} +  e^{j \angle \lambda_{2}} \sqrt[3]{|\lambda_{2}|} - \frac{a}{3},
\end{align}
where $ \angle $ denotes the phase angle of a complex random
variable, and
\begin{align}
\lambda_{1} & = -\frac{q}{2} + \sqrt{\Delta},~\lambda_{2} = -\frac{q}{2} - \sqrt{\Delta}, \nonumber\\
\Delta & = \frac{p^{3}}{27} + \frac{q^{2}}{4},~
p = -\frac{a^{2}}{3} + b,~ q = \frac{2 a^{3}}{27} - \frac{ab}{3}+c,\nonumber\\
a & = 0,~b = \frac{2 A X Y - B X}{Y \|\mathbf{h}\|^{2}},~ c = - \frac{2 A X^{2}}{Y \|\mathbf{h}\|^{2}}.\nonumber
\end{align}	
Thus, the optimal power allocation of the BS can be achieved by substituting \eqref{eq:Optimal_lambda} in \eqref{eq:Optimal_P_BS_no_energy_trading} as
\begin{align}\label{eq:No_energy_trading_P_{BS}_2}
P_{BS}^{\textrm{opt}} = \bigg[ \frac{\mu (1 - \theta)}{\lambda^{\textrm{opt}} \theta \|\mathbf{h}\|^{2}} - \frac{(1 - \theta)\sigma_{s}^{2}}{\xi \theta \|\mathbf{h}\|^{2}|h_{s}|^{2}} \bigg]^{+}.
\end{align}
Noted that for a fixed $ \theta $, both $ U_{BS} $ and $ U_{L} $ can be shown to be the concave functions in terms of $ \lambda $ and $ P_{BS} $, respectively. Thus, we have completed the derivations of the \emph{Stackelberg} equilibrium ($ \lambda^{\textrm{opt}} $, $P_{BS}^{\textrm{opt}}$) for the formulated \emph{Stackelberg} game which are shown in \eqref{eq:Optimal_lambda} and \eqref{eq:No_energy_trading_P_{BS}_2}, respectively.

\subsection{Social Welfare Optimization}
In the previous subsection, we solve the \emph{Stackelberg} equilibrium of the proposed game with energy trading. In order to show the energy price loss of the D2D transmitter due to the self-interested behaviors of the hybrid BS in the proposed game-theoretical scheme, we investigate a social welfare optimization scheme in this subsection. Specifically, we consider that the cooperation between the D2D transmitter and the BS aims to maximize the social welfare, which is defined as the difference between the benefits obtained from the achievable secrecy throughput at the D2D transmitter and the cost of the BS. In this scenario, the energy transfer price will not be considered.
The social welfare optimization is done by jointly optimizing the energy transfer time allocation and the transmit power of the BS. Mathematically, the social welfare utility function can be formulated as \\
\textbf{Social Welfare}
\begin{align}
& U_{SW}(\theta,P_{BS},p_{s},\rho_{e})  = \mu T_{s} - \theta (AP_{BS}^{2} + BP_{BS}) \nonumber\\
&~~~~~~ = \mu(1-\theta) \bigg[ \log\bigg( 1 + \frac{p_{s} |h_{s}|^{2}}{ \sigma_{s}^{2}} \bigg) \nonumber\\&~~~~~~ - \log (1 + \rho_{e})\bigg] - \theta (A P_{BS}^{2} + B P_{BS}).
\end{align}
Thus, the social welfare maximization problem\footnote{This social welfare scheme is considered as an alternative scenario where the D2D transmitter and the BS are deployed and operated by the same service provider.} is given by
\begin{align}\label{eq:Social_welfare_utility}
\max_{\theta,P_{BS},p_{s},\rho_{e}}&~ U_{SW}(\theta,P_{BS},p_{s},\rho_{e}), \nonumber\\
s.t.& ~p_{out} \! \leq \! \varepsilon, ~\rho_{e} \! \geq \! 0,~ 0 \! \leq \! p_{s} \! \leq \! p_{s}^{\max},~ 0 \! < \! \theta \! < \! 1.
\end{align}
The problem \eqref{eq:Social_welfare_utility} is not convex in terms of utility function. To solve this problem, we consider the following steps:
\begin{enumerate}
	\item We first obtain the optimal solutions i.e., $ p_{s}^{\textrm{opt}} $ and $ \rho_{e}^{\textrm{opt}} $ via \emph{Theorem} \ref{theorem Optmal_p_s_and_rho_e} shown in Section \ref{section Energy_trading_stackelberg_game}.
	\item Then, the optimal solution $ P_{BS}^{\textrm{opt}} $ is achieved by solving the problem \eqref{eq:Social_welfare_utility} for given $ p_{s} $ and $ \rho_{e} $.
	\item Finally, the optimal energy time allocation $ \theta^{\textrm{opt}} $ can be achieved via the numerical search shown in Section \ref{section Energy_trading_stackelberg_game}.
\end{enumerate}
In this subsection, we focus on the second step to obtain the optimal power allocation of the BS. First, we fix $ p_{s} $ and $ \rho_{s} $ from Section \ref{section Energy_trading_stackelberg_game}, then the utility function of social welfare can be rewritten as
\begin{align}
U_{SW}(\theta,P_{BS}) &~ = \mu(1-\theta)\bigg[ \log\bigg(1 + \frac{ \theta \xi  P_{BS} \|\mathbf{h}\|^{2}|h_{s}|^{2}}{(1-\theta)\sigma_{s}^{2}}\bigg) \nonumber\\ &~~~~ - \log(1\!+\!\rho_{e}^{\textrm{opt}})\bigg] \!-\! \theta (A P_{BS}^{2} \!+\! B P_{BS}).
\end{align}
The above utility function is concave with respect to $ P_{BS} $ since its second-order derivatives is less than zeros.
Then, take the first-order derivative of the objective function to \eqref{eq:Social_welfare_utility} with respect to $ P_{BS} $ for a given $ \theta $, and equal to zero as follows,
\begin{align}\label{eq:Partial_dervatives_U_SW}
\frac{\partial U_{SW}}{\partial P_{BS}} =  \frac{a d }{1 + d P_{BS}} - 2 \theta A P_{BS} - \theta B = 0,
\end{align}
where $ a = \mu (1-\theta) $, and $ d = \frac{\theta \xi \|\mathbf{h}\|^{2}|h_{s}|^{2}}{(1-\theta)\sigma_{s}^{2}} $.
After some mathematical manipulations, \eqref{eq:Partial_dervatives_U_SW} can be simplified as follows:
\begin{align}
2\theta A d P_{BS}^{2} + (2 \theta A + d B \theta) P_{BS} + (B \theta - a d) = 0.
\end{align}
The optimal power allocation for the hybrid BS can be expressed as
\begin{align}
P_{BS}^{\textrm{opt}} \!=\! \left[\frac{-(2 \theta A \!+\! d B \theta) \!+\! \sqrt{(2 \theta A \!+\! d B \theta)^{2} \!-\! 8 \theta A d (B \theta \!-\! a d)}}{4 \theta A d} \right]^{+}\!\!.
\end{align}
\section{Numerical Results}\label{section Numerical_results}
In this section, we provide simulation results to validate the theoretical results for our proposed \emph{Stackelberg} games. In order to evaluate the performance of these schemes, we consider a secure wireless-powered D2D communication system shown in Fig. \ref{fig:System_model}, which consists of a hybrid BS equipped with five transmit antennas ($ N_{T} = 5 $), a D2D pair, and two eavesdroppers ($ K = 2 $). The maximum allowable secrecy outage probability is $ \varepsilon = 0.1 $, and the energy harvesting efficiency $ \xi = 0.8 $.

\begin{figure}[!htb]
	\centering
\includegraphics[scale = 0.49]{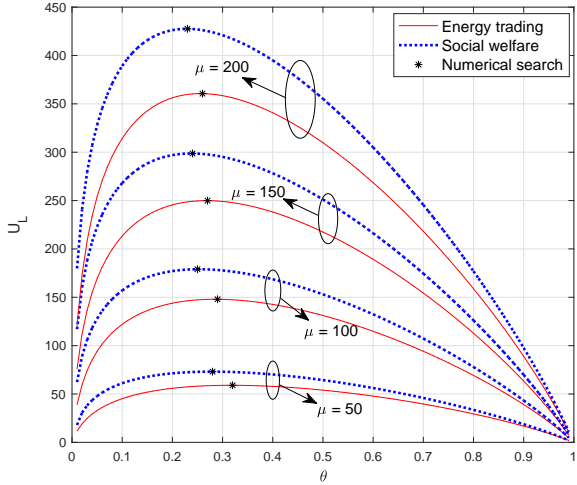}
	\caption{$U_{L}$ versus $ \theta $.}
	\label{fig:U_L_vs_theta}
\end{figure}

\begin{figure}[!htb]
	\centering
	\includegraphics[scale = 0.47]{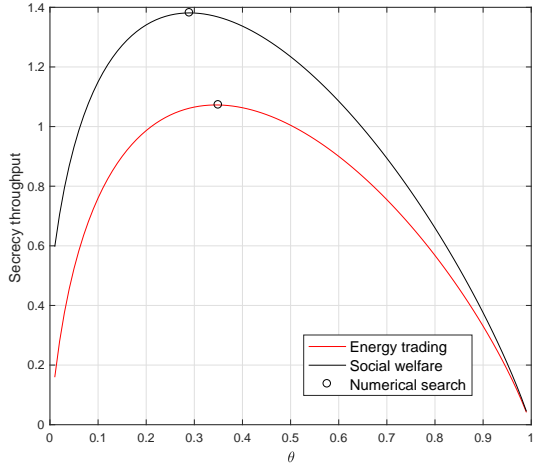}
	\caption{Secrecy throughput versus $ \theta $.}
	\label{fig:Sec_vs_theta}
\end{figure}

First, we validate the concavity of the energy trading based utility function of the D2D transmitter shown in \eqref{eq:U_L_with_theta} with respect to the energy transfer time allocation $ \theta $. Fig. \ref{fig:U_L_vs_theta} shows the utility function versus the energy time allocation $ \theta $ for both energy trading and social welfare schemes. Additionally, a numerical search is considered to obtain the optimal energy transfer time allocation $ \theta^{\textrm{opt}} .$ It is observed that their utility functions are concave in term of $ \theta $ with an optimal energy price $ \lambda^{\textrm{opt}}.$ Moreover, the social welfare optimization scheme achieves a higher utility value because two parties work together for the social welfare purpose. This reflects the fact that the hybrid BS is altruism for wireless energy transfer in the social welfare scheme, whereas it is egoism (or self-interested) in the energy trading based game, which has a direct impact on the D2D transmitter. The same behaviours can also be observed in Fig. \ref{fig:Sec_vs_theta} where the secrecy throughputs for both energy trading and social welfare schemes are shown versus the energy time allocation $ \theta $.

\begin{figure}[!htb]
	\centering
	\includegraphics[scale = 0.49]{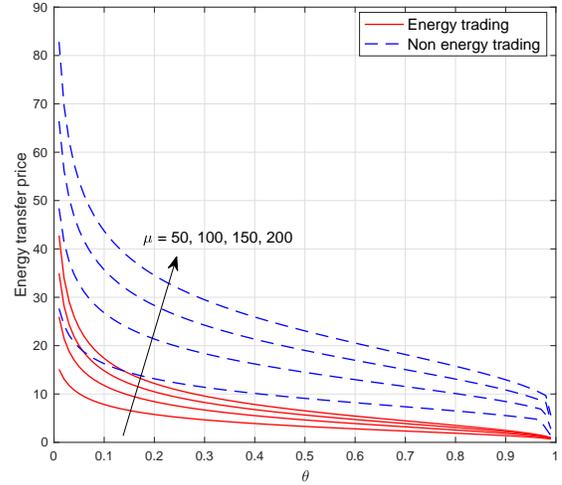}
	\caption{Energy transfer price versus $ \theta $.}
	\label{fig:Price_vs_theta_mu_new}
\end{figure}

Next, the energy transfer price paid by the D2D transmitter to the hybrid BS is evaluated for both the energy trading and non-energy trading schemes in Fig. \ref{fig:Price_vs_theta_mu_new}. From this result, we can observe that the energy transfer price of both energy trading and non-energy trading schemes is a decreasing function of $ \theta.$ This is because of the fact that in this energy interaction with a given time resource of $T$, if more time is spent on energy harvesting, i.e., larger $\theta$, then less time is spent on information transfer, i.e., smaller $(1-\theta)$. Given that the same energy service is purchased by the D2D network, the larger $ \theta $ value would then lead to the lower price to be paid by the D2D transmitter. In addition, it can also be observed that non-energy trading scheme needs to pay a higher price than the energy trading scheme, which highlights the advantage of the energy trading scheme where the customer can negotiate and decide the optimal trading price.

\begin{figure}[!htb]
	\centering
	\includegraphics[scale = 0.47]{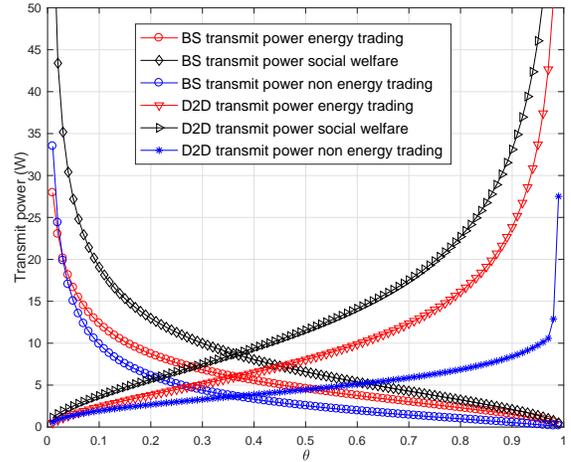}
	\caption{Transmit power ($ P_{BS} $/$ p_{s} $) versus $ \theta $.}
	\label{fig:Power_vs_theta_mu100}
\end{figure}

We also evaluate the transmit power performance, i.e., the BS's and the D2D transmit power, for the three schemes. Fig. \ref{fig:Power_vs_theta_mu100} plots the BS/D2D transmit power, i.e., $ P_{BS} $ or $ p_{s} $, versus $ \theta $. 
It shows that the BS's transmit power, i.e., $ P_{BS} $, for these three schemes decreases as $ \theta $ increases, whereas the D2D's transmit power, i.e., $ p_{s} $, increases with $ \theta $. This implies that the BS requires a lower level of power to support the wireless charging for the D2D transmitter when the WET time allocation $\theta$ increases, while the D2D transmitter would have more power available to support the secure D2D communication as the WIT time allocation $(1-\theta)$ decreases.
It is noted that the D2D transmitter can initially employ the increasing available power to improve the secrecy rate, i.e. increase its utility $U_L$. However, if the D2D's transmit power keeps increasing further as $\theta$ rises beyond the optimal value, c.f. $ \theta^{\textrm{opt}}$ in Fig. \ref{fig:U_L_vs_theta}, the secrecy worsens because the information now becomes more vulnerable to the eavesdropper at high power. This reflects the concave behavior of the utility functions as discussed earlier.

\begin{figure}
	 \centering
	\subfigure[The comparison between the optimal $ \theta $ and fixed $ \theta $.] { \label{fig:U_L_vs_xi_optimal_fix_theta_new}
		\includegraphics[width=0.465\columnwidth]{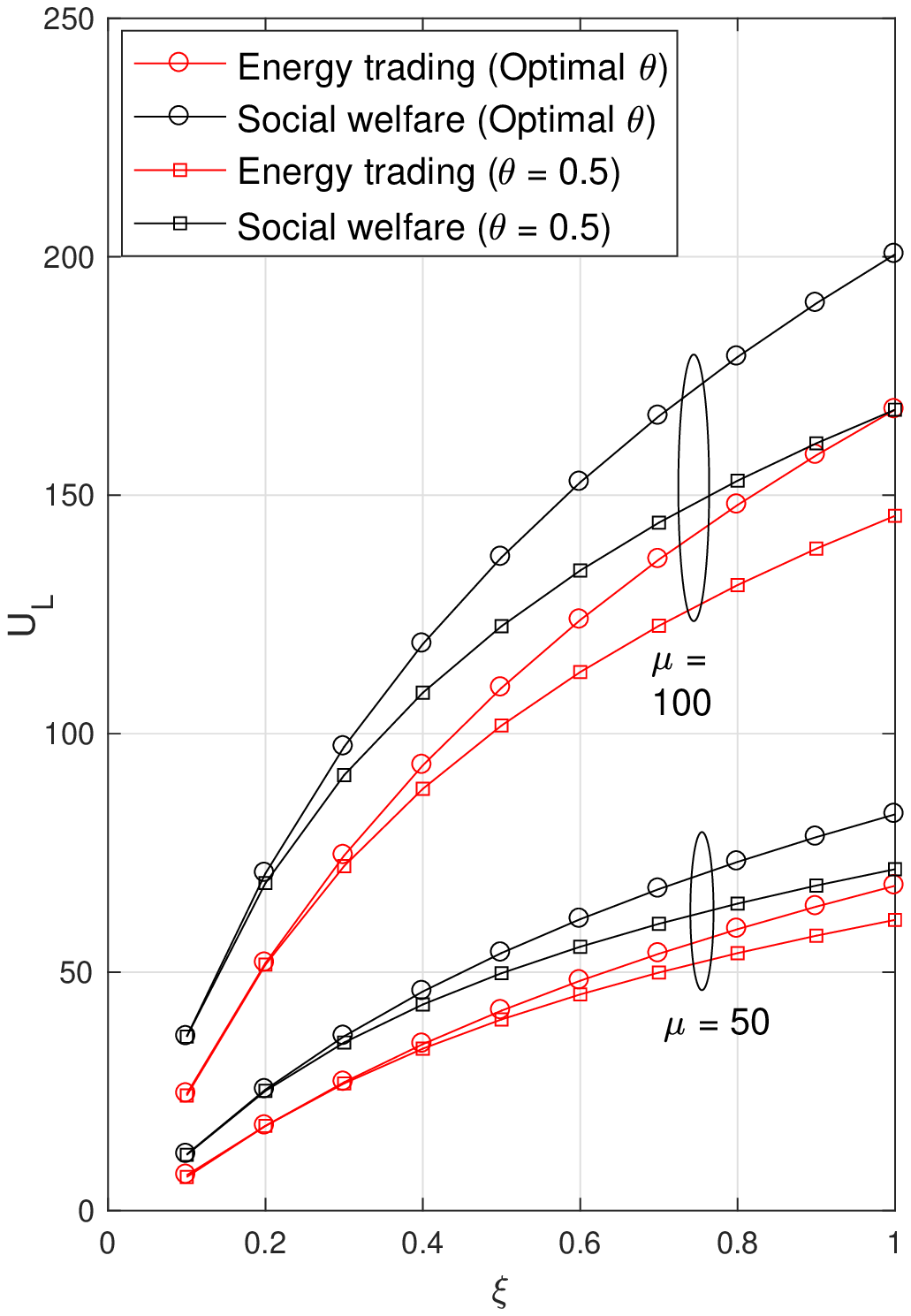} }
	\subfigure[The comparison of three schemes for fixed $ \theta $.] { \label{fig:U_L_vs_xi_fix_mu_theta05}
		\includegraphics[width=0.465\columnwidth]{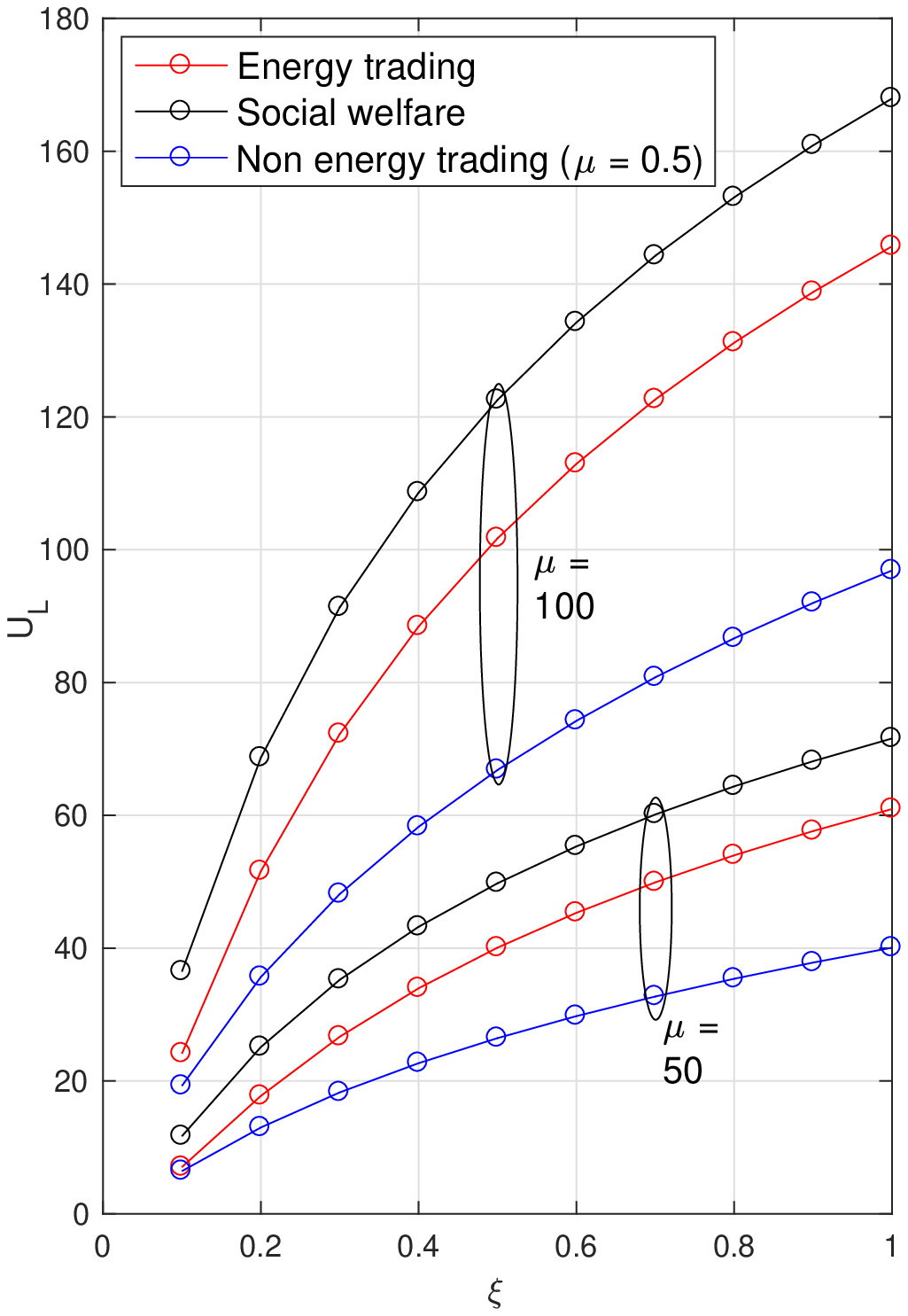} }
	\caption{ $ U_{L} $ vs $ \xi $. }
	\label{fig:U_L_vs_xi}
\end{figure}

Fig. \ref{fig:U_L_vs_xi} shows the utility function versus EH efficiency $ \xi $, where the utility functions increase with $ \xi $. Fig. \ref{fig:U_L_vs_xi_optimal_fix_theta_new} shows that the social welfare optimization scheme outperforms the energy trading scheme in terms of utility function.
In addition, our proposed scheme, i.e., $ \theta^{\textrm{opt}} $, has a higher utility than the fixed $ \theta $ based scheme.
In Fig. \ref{fig:U_L_vs_xi_fix_mu_theta05}, we compare both energy trading and social welfare schemes against the non-energy trading scheme with $ \theta = 0.5 $. As seen in Fig. \ref{fig:U_L_vs_xi_fix_mu_theta05}, both energy trading and social welfare optimization schemes outperform the non-energy trading scheme in terms of D2D user' utility function as the EH efficiency  $ \xi $ increases. This is because of the fact that the D2D users cannot negotiate the price nor can they dictate the energy market when the leader BS decide every things in the first place, which can happen in a monopoly situation when there is only one energy seller.

\begin{figure}
	\centering
	\subfigure[$ \mu = 50 $.] { \label{fig:P_BS_vs_xi_mu50}
		\includegraphics[width=0.465\columnwidth]{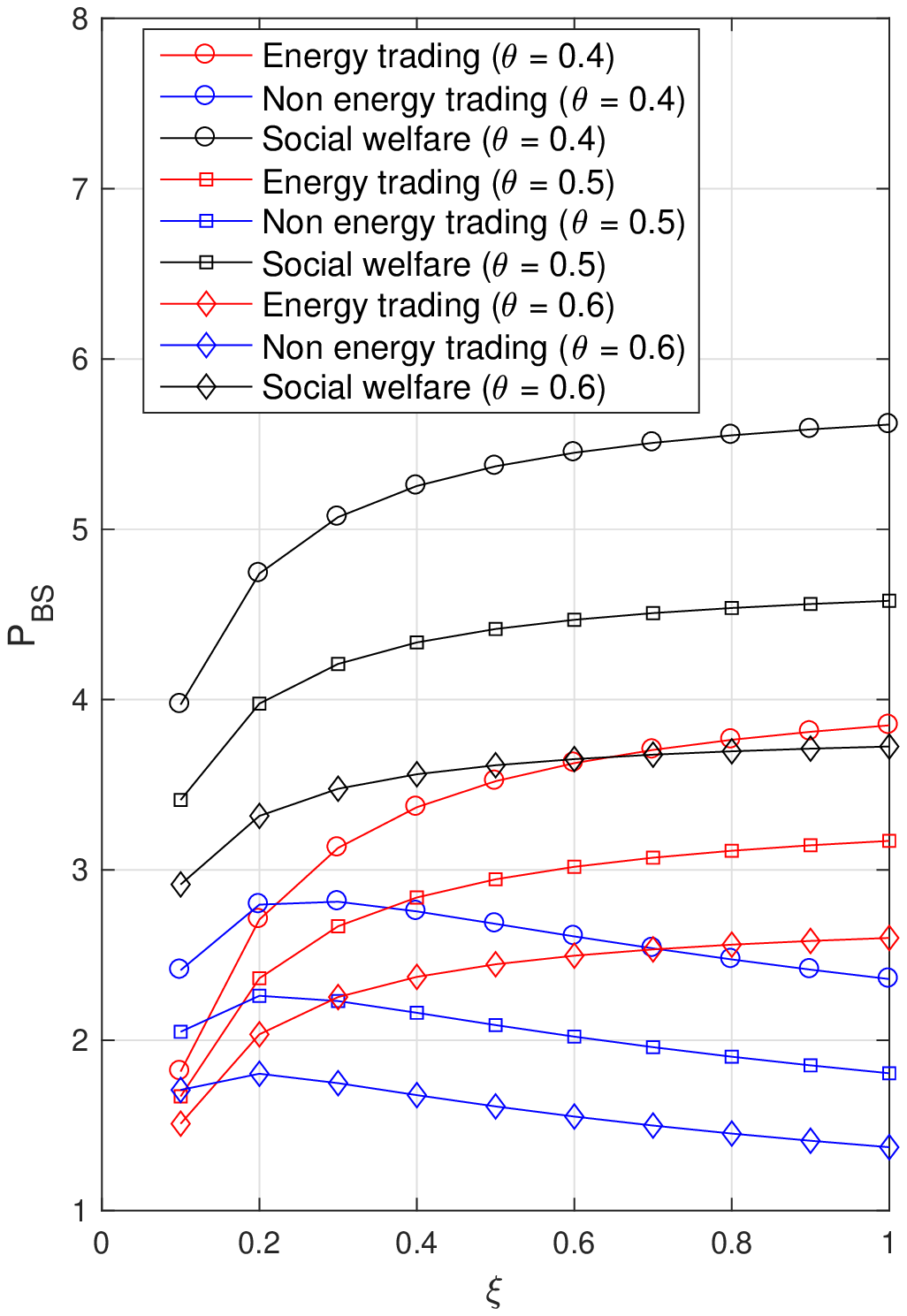} }
	\subfigure[$ \mu = 100 $.] { \label{fig:P_BS_vs_xi_mu100}
		\includegraphics[width=0.465\columnwidth]{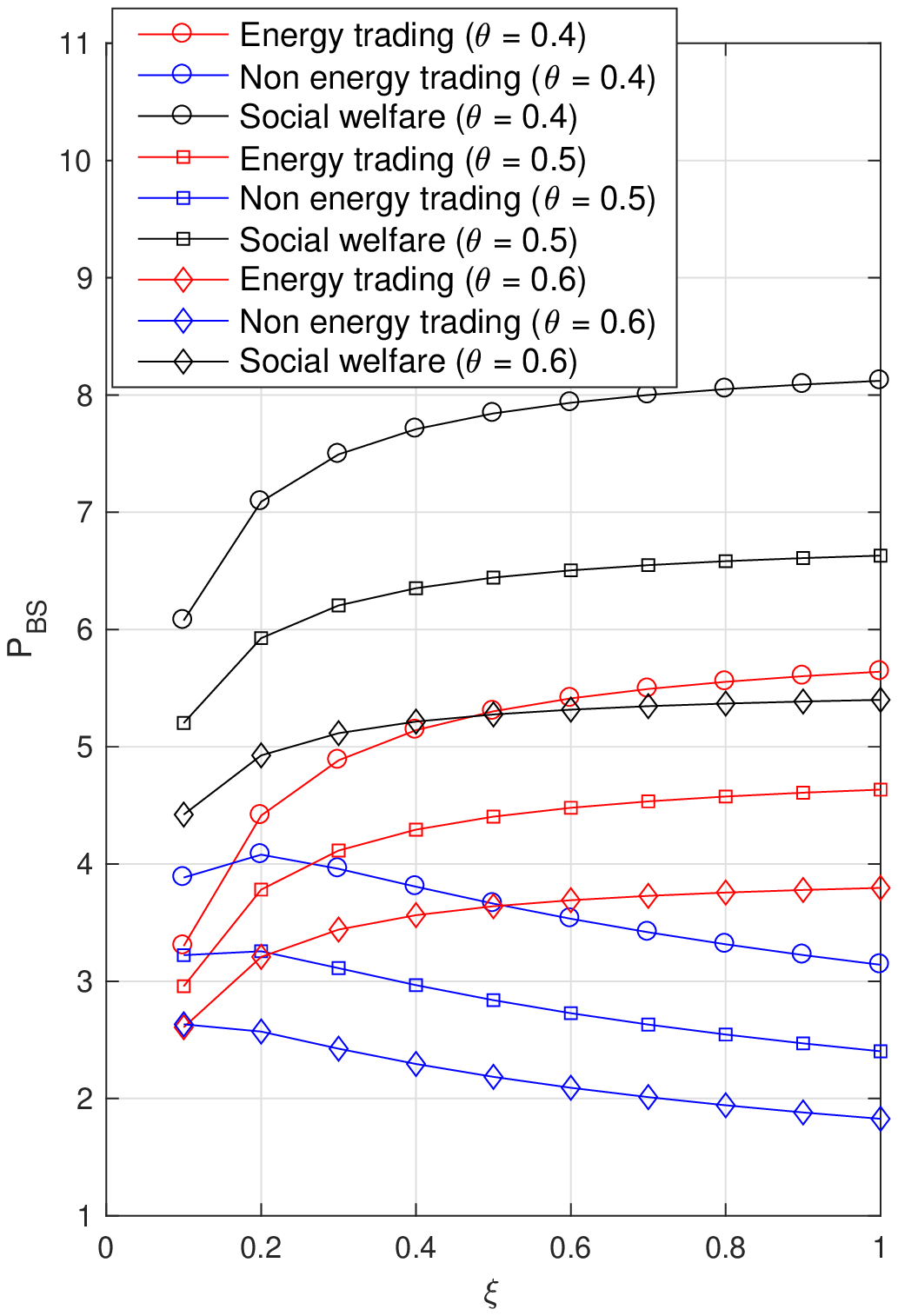} }
	\caption{$ P_{BS} $ vs $ \xi $.}
	\label{fig:P_BS_vs_xi}
\end{figure}

Fig. \ref{fig:P_BS_vs_xi}  shows the hybrid BS transmit power $ P_{BS} $ versus the EH efficiency $ \xi $ for the proposed schemes. It is clear from this figure that both energy trading and social welfare schemes increases with $ \xi $ in terms of the BS transmit power ($ P_{BS} $) and asymptotically stable in the high EH efficiency regime, whereas the non-energy trading scheme first increases with $ \xi $ and then decreases after $ \xi \approx 0.2 $ in terms of the BS transmit power. In addition, the BS transmits more power in non-energy trading scheme than energy trading scheme in the low EH efficiency regime. However, this trend is reversed as $ \xi $  increases in the high EH efficiency regime. This is because of the fact that the non-energy trading scheme needs more power to support the D2D transmitter to facilitate its secure communications in the low EH efficiency regime. Moreover, Fig. \ref{fig:P_BS_vs_xi_mu50} and Fig. \ref{fig:P_BS_vs_xi_mu100} also show that the larger $ \mu $, the more power the BS releases to support the D2D secure transmissions.

\begin{figure}[!htb]
	\centering
	\includegraphics[scale = 0.47]{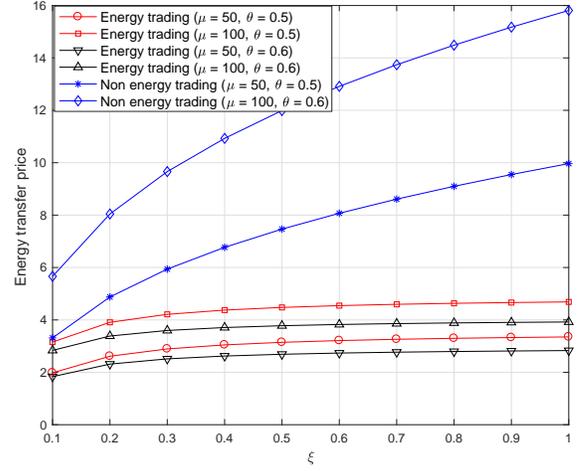}
	\caption{Energy transfer price versus $ \xi $.}
	\label{fig:Price_vs_xi_theta0506}
\end{figure}

Fig. \ref{fig:Price_vs_xi_theta0506} shows the energy transfer price paid by the D2D transmitter with different $ \xi $. It can be observed  from this figure that the energy trading scheme pays less energy price than the non-energy trading scheme, which indicates that the energy trading interaction between the BS and the D2D transmitter can play a cost-efficiency role. In addition, for the energy trading scheme, the larger $ \theta $ is required, the less energy price the D2D transmitter pays.

\begin{figure}[!htb]
	\centering
	\includegraphics[scale = 0.49]{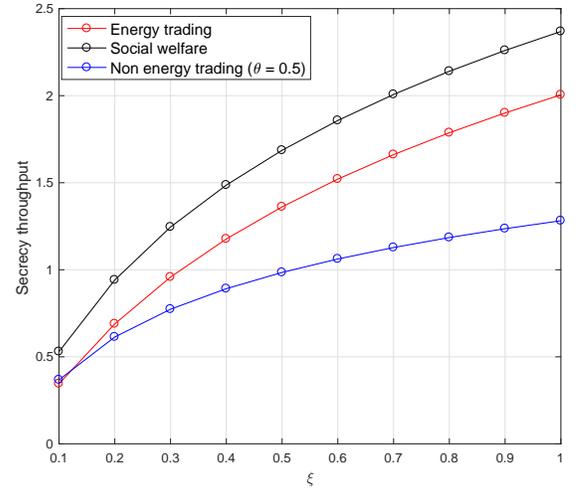}
	\caption{Secrecy throughput versus $ \xi $.}
	\label{fig:Sec_vs_xi}
\end{figure}

Fig. \ref{fig:Sec_vs_xi} illustrates the secrecy throughput performance versus the energy harvesting efficiency $ \xi $. It is clear from this figure that the secrecy throughput increases with $ \xi $, which implies that the secrecy transmission between the D2D pair is guaranteed by improving the energy harvesting efficiency. In addition, both social welfare and energy trading schemes outperform the non energy trading scheme in terms of secrecy throughput, which highlight the advantage of the energy trading interaction between the WET and the D2D network.

\begin{figure}[!htb]
	\centering
	\includegraphics[scale = 0.47]{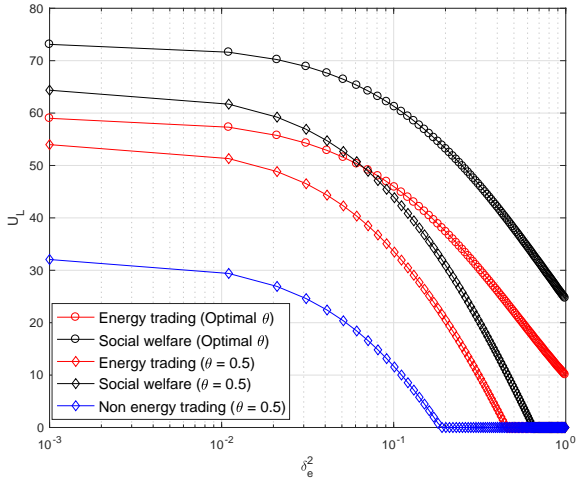}
	\caption{$U_{L}$ versus $ \delta_{e}^{2} $.}
	\label{fig:U_L_vs_Variance_bs_ek}
\end{figure}

\begin{figure}[!htb]
	\centering
	\includegraphics[scale = 0.51]{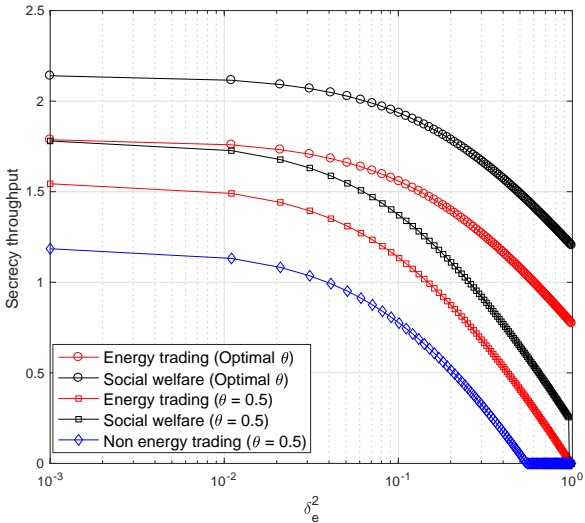}
	\caption{Secrecy throughput versus $ \delta_{e}^{2} $.}
	\label{fig:Sec_vs_delta_e}
\end{figure}

In Fig. \ref{fig:U_L_vs_Variance_bs_ek}, we evaluate the utility function with different jamming signal variances $ \delta_{e}^{2} $. It is shown that the utility function for all the proposed schemes decreases monotonically as $ \delta_{e}^{2} $ increases. In addition, the social welfare scheme outperforms the energy trading scheme.
Moreover, we compare both optimal schemes with the fixed $ \theta $ scheme. The results indicate that the optimal schemes have higher utility functions than the schemes with $ \theta = 0.5 $. It can be observed that both the energy trading and social welfare schemes outperforms the non-energy trading scheme in terms of utility function when $ \theta = 0.5 $. This result confirms that the social welfare scheme gives a better performance because all parties work together in a social corporate responsibility manner. Interestingly, similar behaviours of the proposed schemes can also be seen from Fig. \ref{fig:Sec_vs_delta_e} where we evaluate the secrecy performance with different jamming signal variances $ \delta_{e}^{2} $.

\begin{figure}[!htb]
	\centering
	\includegraphics[scale = 0.44]{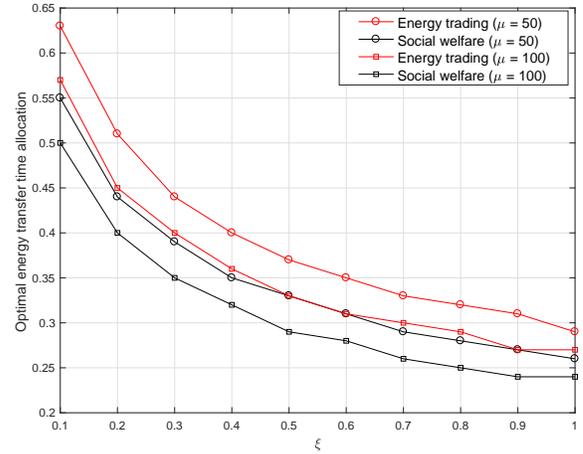}
	\caption{Optimal energy time allocation $ \theta^{\textrm{opt}} $ versus $ \xi $.}
	\label{fig:Optimal_energy_time_allocation_vs_xi}
\end{figure}
Then, we evaluate the WET time allocation with the EH efficiency $ \xi $ in Fig. \ref{fig:Optimal_energy_time_allocation_vs_xi}. It can be observed that the optimal WET time allocation monotonically decreases with $ \xi $. This is due to a fact that the WET phase requires less time to transfer power as the EH efficiency $ \xi $ increases. It is clear that social welfare scheme outperforms the energy trading scheme in terms of the optimal WET time allocation. Furthermore, the larger $ \mu $ is, the smaller the WET time is required.

\begin{figure}[!htb]
	\centering
	\includegraphics[scale = 0.49]{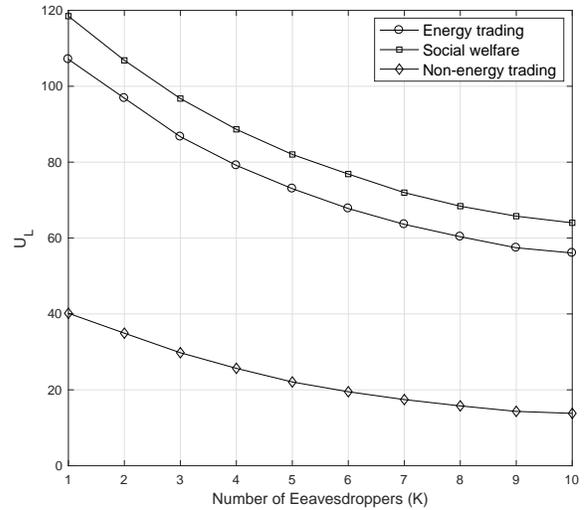}
	\caption{$ U_L $ versus the number of eavesdroppers $ K $.}
	\label{fig:U_L_vs_eves}
\end{figure}

Finally, the utility function versus the number of the eavesdropper $ K $ is evaluated in Fig. \ref{fig:U_L_vs_eves}. From this result, it is observed that the utility function decreases as $ K $ increases. In addition, we compare the three proposed schemes where the social welfare scheme outperforms the energy trading scheme. The non-energy trading scheme demonstrates the worst utility among these three schemes, which implies that the non-energy trading scheme does not provide any advantage for the D2D users when they act as a passive customer in an environment where the leader BS can decide the energy price in a monopoly manner, particularly when the number of eavesdroppers are large.
 
\section{Conclusion}\label{section:Conclusions}
This paper investigated a secure wireless-powered CJ-aided D2D communication system, where a hybrid BS in a cellular network provides power wirelessly for the D2D transmitter and at the same time plays a CJ’s role to interfere with the eavesdroppers to support the secure D2D communications. We have formulated different scenarios and solved different problems of energy interactions using the \textit{Stackeberg} game. We found that a higher utility level is achieved if social welfare responsibility is taken into account. This is understandable, as all parties work together to maximize a common target. At the same time, the energy buyer, i.e., the D2D user, would have to pay a higher price for energy purchase if the seller, i.e., the BS, has a monopoly authority to dictate and lead the market, as shown in the non-energy trading scenario. However, in an environment where there are potential competitors for energy selling, the D2D customer can become the leader who would negotiate for a much better energy price, as reflected in the energy trading case of our proposed game.
In addition, numerical results were provided to validate our proposed schemes and showed that both energy trading and social welfare schemes provide a better energy cost efficiency. Overall, the results highlighted the importance of the energy trading interactions between the cellular and D2D networks.
\bibliographystyle{ieeetr}
\bibliography{my_references}
\begin{IEEEbiography}{Zheng Chu} (M'17) is with 5G Innovation Center (5GIC), Institute of Communication Systems (ICS), University of Surrey, U.K.. He was with the Faculty of Science and Technology, Middlesex University, London, U.K. from Sept. 2016 to Oct. 2017. Prior to this, he received Ph.D. degree in School of Electrical and Electronic Engineering, Newcastle University, U.K., in 2016. His research interests include physical layer security, wireless cooperative networks, wireless power transfer, convex optimization techniques, and game theory.
\end{IEEEbiography}

\begin{IEEEbiography}{Huan X. Nguyen} (M'06-SM'15) received the B.Sc. degree with the Hanoi University of Science and Technology, Vietnam, in 2000, and the Ph.D. degree from the University of New South Wales, Australia, in 2007. He has since been with several universities in the U.K (Research Officer at Swansea University during 2007-2008 and Lecturer at Glasgow Caledonian University, 2008-2010). He is currently an Associate Professor of Communication Networks at the Faculty of Science and Technology, Middlesex University, London, U.K. His research interests include 5G enabling technologies, PHY security, energy harvesting, and communication systems for critical applications. He has published more than 90 research papers, mainly in the IEEE journals and conferences. He received a grant from the Newton Fund/British Council Institutional Links program (2016-2018) for Disaster Communication and Management Systems using 5G Networks. He was the co-chair of the 2017 International Workshop on 5G Networks for Public Safety and Disaster Management (IWNPD 2017).  Prof. Nguyen is a Senior Member of the IEEE. He is currently serving as the Editor of the KSII Transactions on Internet and Information Systems.
\end{IEEEbiography}

\begin{IEEEbiography}{Tuan Anh Le} (S'10-M'13) received his B.Eng. and M.Sc. degrees both in electronics and telecommunications from Hanoi University of Technology, Hanoi, Vietnam, in 2002 and 2004, respectively, and his Ph.D. degree in telecommunications research from King's College London, The University of London, U.K., in 2012.  From 2009 to 2012, he was a researcher on the Green Radio project funded by the Core 5 joint research program of the U.K.'s Engineering and Physical Sciences Research Council (EPSRC) and the Virtual Center of Excellence in Mobile and Personal Communications (Mobile VCE).  From July 2013 to October 2014, he was a Post-Doctoral Research Fellow within the School of Electronic and Electrical Engineering, University of Leeds, Leeds, U.K. Since November 2014, he has been a lecturer with the Faculty of Science and Technology, Middlesex University, London, U.K. His current research interests are cooperative communications, D2D communications, cognitive radio, RF energy harvesting and wireless power transfer, physical-layer security, robust resource allocation and interference management in 5G cellular networks, and channel estimation and resource allocation techniques for massive MIMO. He was a recipient of the prestigious Ph.D. scholarship jointly awarded by the Mobile VCE and the U.K. Government’s EPSRC.
\end{IEEEbiography}

\begin{IEEEbiography}{Mehmet Karamanoglu} is Professor in Design Engineering at Middlesex University.
		He graduated with a BEng degree in Mechanical Engineering and followed onto to
		complete his PhD in numerical methods, supported by British Aerospace. He is
		currently heading the department of Design Engineering and Mathematics. His
		expertise includes manufacturing automation, CAD, design engineering, modelling
		and robotics. His research interests include numerical analysis, process simulation,
		design strategies and robotic systems.
\end{IEEEbiography}

\begin{IEEEbiography}{Enver Ever} (M'13) obtained his BSc. Degree from the Department of Computer Engineering, Eastern Mediterranean University, Cyprus in 2002. He then continued his studies at Middlesex University where he obtained his MSc in Computer Networks and PhD in Performance Evaluation of Computer Networks and Communication Systems in 2004 and 2008 respectively. He worked at Bradford University as a postdoctoral Research Associate for a year. Following that he worked as a Senior Lecturer in Computer and Communications Engineering Department Middlesex University. Currently he is an Associate Professor in Middle East Technical University, Northern Cyprus Campus. His current research interests include computer networks, wireless communication systems, public safety networks, internet of things, cloud computing, wireless sensor networks, and performance/reliability modelling. He serves on various Programme Committees and received exemplary reviewer award for his contributions as reviewer. 
\end{IEEEbiography}

\begin{IEEEbiography}{Adnan Yazici} (SM'99) is a full professor and the chair of Department of Computer Engineering at Middle East Technical University, Ankara-Turkey. He received his PhD in Computer Science from the Department of EECS at Tulane University in USA, in 1991. His current research interests include; intelligent database systems, multimedia and video databases and information retrieval, wireless multimedia sensor networks, data science, and fuzzy database modeling. He received the IBM Faculty Award in 2011 and the Parlar Foundation's Young Investigator Award in 2001. He has published more than 200 international technical papers and co-authored/edited three books, which are titled as Fuzzy Database Modeling, Fuzzy Logic in Its 50. Year: New Developments, Directions and Challenges, and Uncertainty Approaches for Spatial Data Modeling and Processing: a Decision Support Perspective (all by Springer). Prof. Yazici has been a Conference Co-Chair of the 23rd IEEE International Conference on Data Engineering (ICDE-2007), the 38th Very Large Data Bases (VLDB 2012) and the 23rd IEEE International Conference on Fuzzy Systems (FUZZ-IEEE 2015). He is currently a member of IEEE Computational Intelligence Society (CIS) and Fuzzy Systems Technical Committee (FSTC) and an Associate Editor of IEEE Trans. on Fuzzy Systems. He is also the director of the Multimedia Database Lab. at METU.
\end{IEEEbiography}
\end{document}